# Unveiling contrasting impacts of heat mitigation and adaptation policies on U.S. internal migration

Chao Li[1,2], Xing Su*[3], Chao Fan[4], Yang Li[2], Luping Li[1], Chunmo Zheng[1,5], Wenglong Chao[1], Leena Järvi[2,6], Han Lin[7], Juan Tu[8]

[1] College of Civil Engineering and Architecture, Zhejiang University, Hangzhou, Zhejiang, 310058, China;

[2] Institute for Atmospheric and Earth System Research (INAR)/Physics, Faculty of Science, University of Helsinki, Helsinki, 00014, Finland;

[3] Business School, Hohai University, Nanjing, Jiangsu, 211100, China;

[4] College of Engineering, Computing, and Applied Sciences, Clemson University, Clemson, SC, 29631, USA;

[5] Department of Engineering, Newcastle University, Newcastle NE1 7RU, UK;

[6] Helsinki Institute of Sustainability Science (HELSUS), Faculty of Science, University of Helsinki, Helsinki 00014, Finland;

[7] School of Engineering Audit, Nanjing Audit University, Nanjing 211815, China;

[8] Department of Physics, Nanjing University, Nanjing 210093, China.

## Abstract

While climate-induced population migration has received rising attention, the role played by human climate endeavors remains underexplored. Here, we combine machine learning with attribution mapping to analyze the impacts of 4,713 heat-related policies (HPs) on 11,177 migration flows between U.S. counties. We find that heat adaptation policies (APs) and heat mitigation policies (MPs) have significant and opposing impacts on internal migration: APs reduce out-migration, while MPs increase it. These policies have heterogeneous effects on migration among policy types. Behavioral/cultural MPs at origins lead to a 0.24%–0.68% (95% confidence interval) increase in annual outflows per policy, whereas behavioral/cultural APs at destinations elevate outflows of origins by 0.11%–1.55% (95% confidence interval). Migration patterns are nonlinearly moderated by income, ageing, education, and racial diversity of both origin and destination counties. Ageing rates have the most noticeable U-shaped relationship in shaping migration responses to behavioral/cultural MPs at origins, and inverted U-shapes for institutional MPs at origins and nature-based MPs at destinations. These findings offer critical insights for policymakers on how HPs influence migration as global warming and policy interventions persist.

## Introduction

Anthropogenic warming is intensifying worldwide, resulting in more frequent and severe natural disasters such as heat waves, floods, and hurricanes[1–4]. In response, human climate endeavors to mitigate and adapt to these adverse climatic conditions have increased dramatically in recent years[5–7]. For example, in comparison to 2017, more than 50% of global climate adaptation actions and more than four times as many climate mitigation actions were recorded in the Carbon Disclosure Project (CDP) in 2020 (**Supplementary Fig. S1**).

Despite the increasing simultaneity of climate risks and responses, the influence of climate policies on climate-induced migration remains poorly understood. Although political factors have been broadly recognized as affecting migration dynamics[8,9], numerous empirical studies and meta-analyses have not explicitly isolated the role of climate policies in shaping climate-induced population migration[10–15], partly due to methodological challenges in operationalizing policy variables and data scarcity. Meanwhile, several works have performed numerical simulations to project future migration patterns, often assuming that the implementation of climate policies will reduce out-migration[16–18], yet these assumptions lack robust empirical support. This knowledge gap raises concerns about the accuracy and credibility of both explanatory studies and future projections of climate-driven population movements. Hence, it is imperative to quantify whether and the extent to which, climate policies are influencing migration decisions.

*Corresponding author: Xing Su. Email address: 20250612@hhu.edu.cn

Climate policies affect population migration through a variety of mechanisms that depend on the content of the policy, local economic, social and demographic conditions, etc. Notably, the impact of climate policies on migration is not always the taken-for-granted "pulling effect", but under some conditions, climate policies can push the outflows of population. For example, while heat adaptation policies that build green infrastructure, such as parks, may attract inflows of people by enhancing livability[19], they may also result in green gentrification[20]. Green gentrification refers to the phenomenon that green space leads to an increase in property values, house prices, and thus results in the displacement of the original occupants[21]. Furthermore, the out-migration effect of climate policies comprises not only visible impact pathways but also a large portion of invisible pathways, such as influencing the perceptions and behaviors of the residents. Institutional and behavioral/cultural climate policies exemplify this. For example, to reduce carbon emissions, policies such as carbon taxes and quotas are increasingly being implemented around the world[22]. These policies can be disruptive to the local high-carbon-releasing industries and enterprises, as well as the traditional work and life-style of residents[23].

While the underlying mechanisms through which climate policies shape migration decisions are complex and multifaceted, this study does not aim to decompose each causal pathway. Instead, we focus on empirically estimating the average total effect of climate policies on migration flows, conditional on local and temporal heterogeneity. To identify this effect in the presence of potential confounders—such as time-invariant county characteristics, national policy trends, and unobserved regional shocks—we employ a panel-data strategy that leverages fixed-effects specifications rooted in the gravity-model tradition. This allows us to control for a broad set of confounding influences, even in the absence of full knowledge of all causal pathways (see **Methods** for details).

We take the U.S. as the testbed for our empirical analysis. According to the 2017–2020 CDP climate policy dataset, the U.S. accounted for 27.4% of all climate adaptation policies and 27.0% of mitigation policies globally (**Supplementary Fig. S1**), highlighting its climate ambition of the U.S. as the most populous developed country. Prior research has shown that although climate-migration relationships vary widely in direction and magnitude, slow-onset climate risks such as extreme heat and precipitation generally have a greater migration effect[24]. Furthermore, empirical studies and meta-analyses increasingly find that climate-related drivers are more influential for short-distance migration within countries than for cross-border migration[11,13]. We thus focus on assessing the influence of HPs on internal migration of the contiguous U.S. in the context of extreme heat.

The panel data from the Internal Revenue Service's (IRS) county-to-county migration estimates[25] and the HP dataset from the CDP[26] are employed in this study (see **Methods** for details). To uncover more precise heterogeneous impacts of different kinds of policies, we classify both heat adaptation climate policies (APs) and heat mitigation climate policies (MPs) as technological/ infrastructural, institutional, behavioral/cultural, and nature-based policies referred to the findings of Lea Berrang-Ford et al[7] (**Supplementary Note S1**). Overall, our study encompasses 4,713 HPs for the period 2017-2020, including 768 APs and 3,945 MPs. A considerable number of HPs cannot be reliably categorized by traditional assessment methods, so we trained a supervised machine learning classifier based on the DistilBERT model[27] (see **Methods** for details).

*Corresponding author: Xing Su. Email address: 20250612@hhu.edu.cn

This study addresses three primary questions: 1) Do HPs have a causal impact on internal population migration in the U.S.? 2) If so, what are the heterogeneous effects of this impact in terms of policy types? 3) How do local contextual factors moderate such heterogeneous effects?

To answer the first question, we initially tested the correlation between HP implementation and migration using simple linear regression, Pearson's correlation coefficient, and Spearman's rank correlation coefficient. Acknowledging the limitations of these methods for obtaining causal effects, we introduce the gravity-based fixed-effects models to analyze the bilateral population flows across counties and compute impacts of HPs on migration patterns. To ensure robustness, we conduct extensive sensitivity analyses, including alternative model specifications, different data preprocessing pipelines, and lagged policy effects. In response to the second question, we extend the baseline model to assess type-specific impacts of different HP categories. To tackle the third question, we incorporate socio-economic covariates into the HP-migration dataset and construct both linear and non-linear (quadratic) interaction models to examine how local conditions shape the effects of HPs on migration.

## Evidence on heat-related policies and population migration

Heat-related policies (HPs) for 2017-2020 were widely distributed across 106 origin counties (**Fig.1a**) that implemented an average of 12.02 HPs over the 4-year period (**Supplementary Table S2**), with an average out-migration rate of 1.3% (**Supplementary Table S3**). Detailed lists of origin counties and descriptive statistics for different types of HPs are presented in **Supplementary Tables S1 and S2**, respectively.

Paired t-tests reveal that both heat adaptation policies (APs) and heat mitigation policies (MPs) were significantly more prevalent in destination counties than in origin counties (**Fig. 1b**). On average, destination counties implemented 40.4% more APs and 23.7% more MPs than their origin counterparts (**Supplementary Table S2)**. Across both APs and MPs, technological/infrastructural policies were the most frequently implemented, while nature-based policies were the least common. The spatial distributions of HP subtypes at origin and destination counties are shown in **Supplementary Figs. S2 and S3**, respectively. Regionally, the western U.S. exhibited the highest average number of policies, whereas the northeastern region had the lowest.

Geographic variation in out-migration rates reveals that the South experienced the highest average rate (1.58%), while the Midwest exhibited the lowest (0.88%) (**Fig. 1a**). State-level patterns of absolute migration flows are shown in **Extended Data Fig. 1**. California, Texas, and Florida had the largest out-migrant populations; notably, the majority of these migrants relocated within the same state. This finding is consistent with prior research highlighting the predominance of short-distance, intra-state migration[28].

We conducted statistical analyses using simple linear regressions, Pearson's correlation coefficients and Spearman's rank correlation coefficients (**Fig.1c-f**) to investigate the correlations between HPs and migration. MPs in origin counties exhibited a significant positive association with out-migration across all three methods (**Fig. 1d**). In contrast, APs in origin counties were positively but insignificantly correlated with out-migration (**Fig. 1c**). Detailed correlation coefficients for all HP subtypes are reported in **Supplementary Figs. S4 and S5**. Among these, institutional MPs in origin counties and technological/infrastructural MPs in destination counties showed the strongest and most significant positive correlations with origin-county out-migration rates.

*Corresponding author: Xing Su. Email address: 20250612@hhu.edu.cn

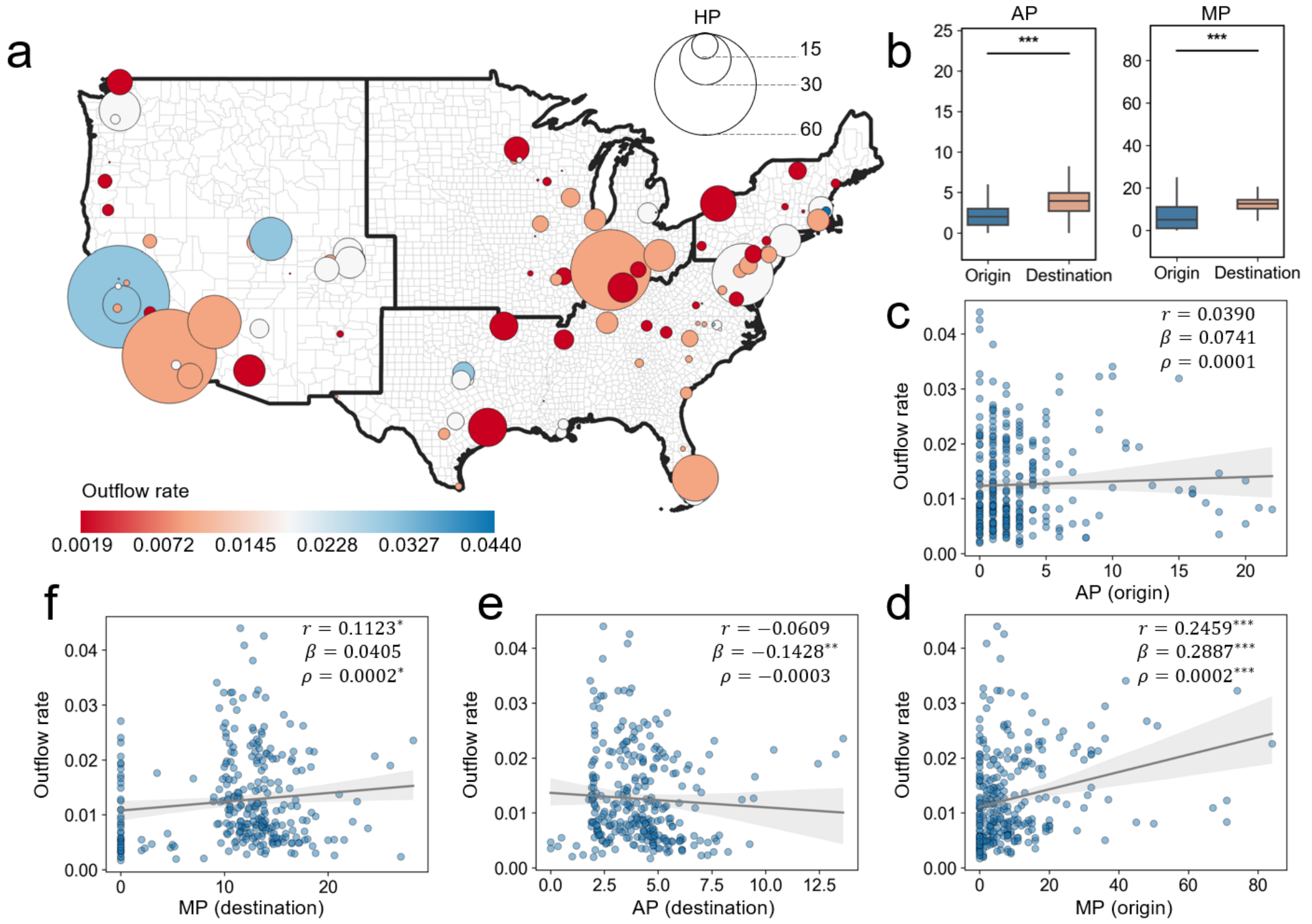

**Fig. 1 Correlations of the heat-related policies (HPs) and the population out-migration. a**, Spatial distribution of heat-related policies (HPs) and outflow rate in 2017-2020. The size and color of the circles in the figure represent the average number of HPs implemented and the average outflow rate of each origin county over the 2017-2020 period, respectively. Black solid lines are U.S. Census 2018 boundary lines for four regions of the continental U.S. Gray solid lines are U.S. County boundary lines, and the base map is from geodatabases of 2019 TIGER/Line shapefiles for US geographical boundaries at the county level (available at https://catalog.data.gov/). **b**, Boxplots of HPs distribution in origin and destination, indicating the 97.5th percentile, upper quartile, median, lower quartile, and 2.5th percentile of the data from top to bottom, respectively. The symbol at the top of the figure indicates the significant $P$ values obtained from paired t-tests for the differences between the two groups. $^{*}P < 0.1$, $^{**}P < 0.05$ and $^{***}P < 0.01$. **c-f**, Scatterplots of outflow rate and HPs. The solid gray line indicates the line of fit for the simple linear regression, and the shaded area indicates the 95% confidence interval. The $r$ values in the figures indicate the Pearson coefficients, the $\beta$ values indicate Spearman's rank correlation coefficients, and $\rho$ indicates the coefficient of the linear regression. The symbol in the upper right corner of the figure indicates the significant $P$ values obtained from t-tests for that coefficient. $^{*}P < 0.1$, $^{**}P < 0.05$ and $^{***}P < 0.01$.

## Causal effects of heat-related policies on population migration

Beyond the preliminary correlation analyses using aggregated data, we further estimate the causal impact of heat-related climate policies (HPs) on population migration by combining bilateral county-level migration flows with policy data in a series of gravity-type fixed effects models.

*Corresponding author: Xing Su. Email address: 20250612@hhu.edu.cn

With the exception of the random effects model in the second column of **Supplementary Table S4**, which itself serves as a reference group, all other models incorporate origin-destination paired fixed effects and year fixed effects. Each model also reports corresponding F-test statistics from pooled OLS regressions and Hausman test statistics from random effects regressions (**Supplementary Table S4**). Both sets of tests support the validity of our gravity-based fixed effects approach.

Results from the main specification show that heat mitigation policies (MPs) implemented at the origin have a significant and positive effect on out-migration: each additional MP is associated with a 0.21% increase in out-migration (±0.0016, 90% CI, $P$ = 0.000; **Fig. 2b, row 1**). In contrast, heat adaptation policies (APs) at the origin are not significantly associated with migration (**Fig. 2a, row 1**). At the destination, each additional MP corresponds to a 0.15% reduction in population outflow from the origin to the destination (±0.0005, 90% CI, $P$ = 0.000; **Fig. 2d, row 1**), while each additional AP is associated with a 0.20% increase in origin-to-destination migration (±0.0016, 90% CI, $P$ = 0.040; **Fig. 2c, row 1**). These findings suggest asymmetric push–pull effects of HPs depending on the policy type and spatial implementation.

To test the robustness of these results, we apply multiple alternative model specifications. Under different data preprocessing strategies (**Fig. 2, rows 3–6**), the magnitude and significance of MP–migration associations remain largely stable (**Fig. 2b and 2d, rows 4–6**), whereas the AP–migration relationships exhibit greater variability, although the direction of the point estimates remains consistent (**Fig. 2a and 2c, rows 4–6**).

We further investigate potential lagged effects of HPs (**Fig. 2, rows 7–8**). The effect of APs at the origin remains statistically insignificant even with time lags (**Fig. 2a, rows 3–8**), while the effect of MPs at both origin and destination also becomes insignificant with lagged specifications (**Fig. 2b and 2d, rows 7 – 8**). This suggests that the migration-inducing effect of MPs occurs contemporaneously and lacks persistence over time. By contrast, APs at the destination exhibit significant lagged effects: both one-year and two-year lags show significance at the 10% level, with effect sizes increasing over time (**Fig. 2c, rows 7–8**). This may indicate a delayed but reinforcing influence of destination-based APs on encouraging migration inflows.

*Corresponding author: Xing Su. Email address: 20250612@hhu.edu.cn

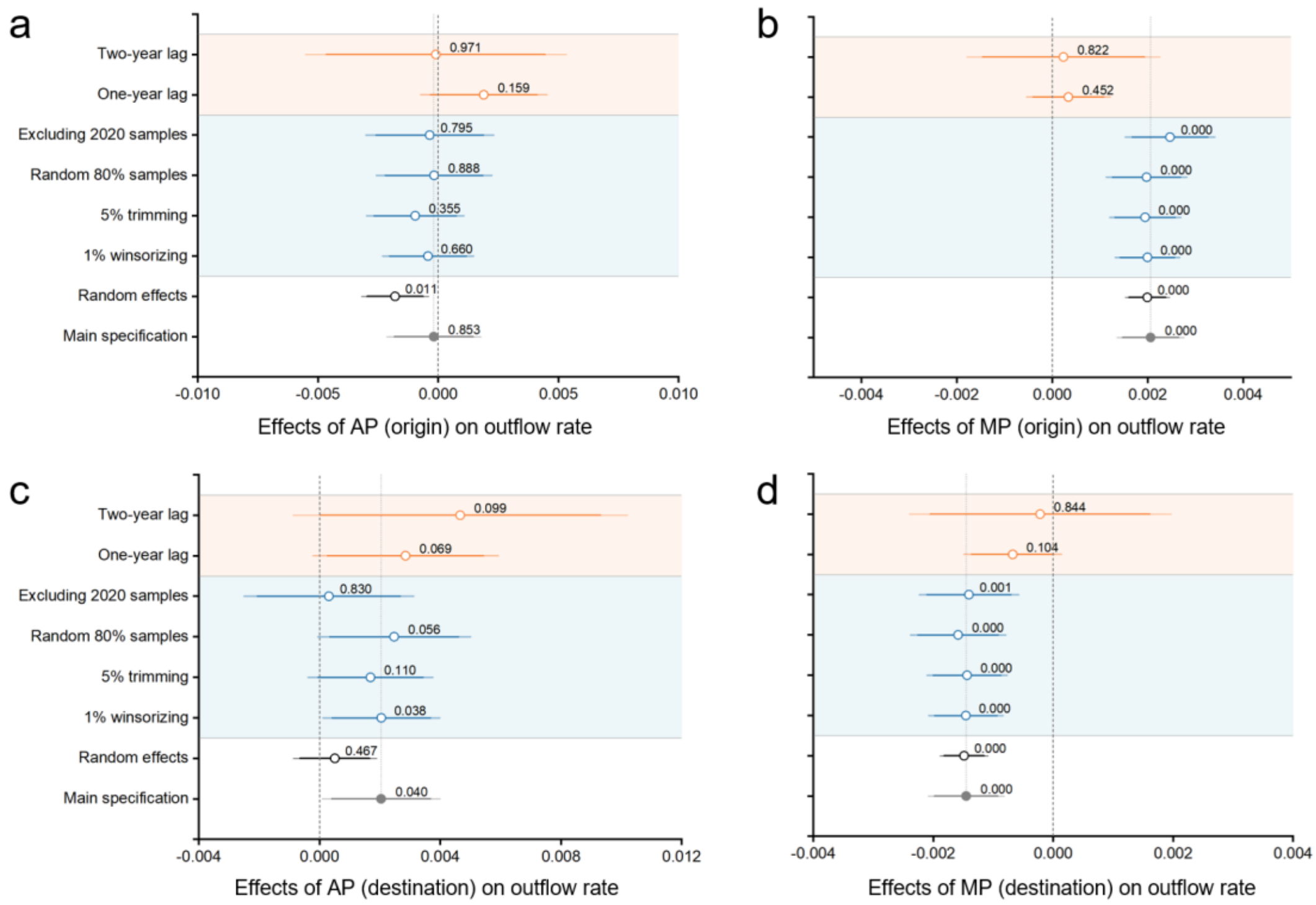

**Fig. 2 Estimated impact on outflow rate of the original county for one additional heat-related policy (HP). a-b**, The impact of AP and MP implemented in the origin county, respectively. **c-d**, The impact of AP and MP implemented in the destination county, respectively. The *x* axis shows the marginal effects and error bars of a 1 increase in policies on out-migration rates. The main specification in the bottom row uses gravity-based fixed effects models via equation (1) ($n$ = 11,177). Other estimates above show the robustness of the main specification to the alternative: a random effects model (row 2, $n$ = 11,177), 1% winsorizing by outflow rate data (row 3, $n$ = 11,177), 5% trimming by outflow rate data (row 4, $n$ = 10,060), using random selected 80% samples of all samples (row 5, $n$ = 8,942), excluding 2020 samples (row 6, $n$ = 7,924), one-year lagged effects (row 7, $n$ = 8,549) and two-year lagged effects (row 8, $n$ = 5,492). In **a-d**, central estimates shown with circles are regression point estimates, neighboring numbers indicate significant $P$ values of estimates computed by t-tests, dark-colored lines indicate 90% confidence intervals and light-colored lines indicate 95% confidence intervals.

## Policy-type heterogeneity in migration effects

To examine how migration responses vary by types of heat-related policies (HPs), we re-estimate the out-migration rate of origin counties using external model specifications. In this part, we still begin by comparing gravity-based fixed effects models with pooled OLS and random effects models. Results confirm the superior and stable performance of gravity-based models (**Table 1, model (1)**).

For HPs implemented at origin counties, we find significant positive effects of institutional and behavioral/cultural heat mitigation policies (MPs) on out-migration (**Fig. 3**). Specifically, each additional institutional MP is associated with a 0.37% increase in the out-migration rate (±0.0020, 90% CI, $P$ = 0.002), while each additional behavioral/cultural MP leads to a 0.46% increase (± 0.0018, 90% CI, $P$ = 0.000).

For HPs implemented at destination counties, more varied and statistically significant effects are observed. One additional behavioral/cultural heat adaptation policy (AP) increases the out-migration rate from the origin by 0.83% (±0.0061, 90% CI, $P$ = 0.025). In contrast, one additional technological/infrastructural MP at the destination leads to a 0.16% decrease (±0.0015, 90% CI, $P$ = 0.084) in out-migration, while one additional behavioral/cultural MP results in a 0.35% decrease (±0.0019, 90% CI, $P$ = 0.002). Interestingly, nature-based MPs at the destination appear to encourage migration, with each additional policy associated with a 0.82% increase in origin-to-destination migration (±0.0066, 90% CI, $P$ = 0.040).

*Corresponding author: Xing Su. Email address: 20250612@hhu.edu.cn

Although earlier results (**Fig. 2**) suggest that the effects of APs and MPs on migration tend to be contemporaneous, we further test for temporal heterogeneity across different policy types to provide a more comprehensive understanding (**Table 1, models (2) and (3)**). The time dynamics of these heterogeneous effects vary substantially across policy categories. For instance, the effect of technological/infrastructural MPs at the destination intensifies over time. In the implementation year, such policies are associated with a 0.16% decrease in out-migration. However, with a one-year lag, the effect reverses, producing a 0.47% increase in out-migration (±0.0027, 90% CI, $P$ = 0.004), which further increases to 0.62% with a two-year lag (±0.0053, 90% CI, $P$ = 0.056). This may indicate that technological and infrastructural investments gradually become more attractive or influential in shaping migration decisions. Conversely, the effect of behavioral/cultural MPs at the destination shows a diminishing trend over time. These policies initially reduce out-migration from the origin, but their influence weakens, suggesting that behavioral or cultural shifts may act as increasingly important push factors in the long run.

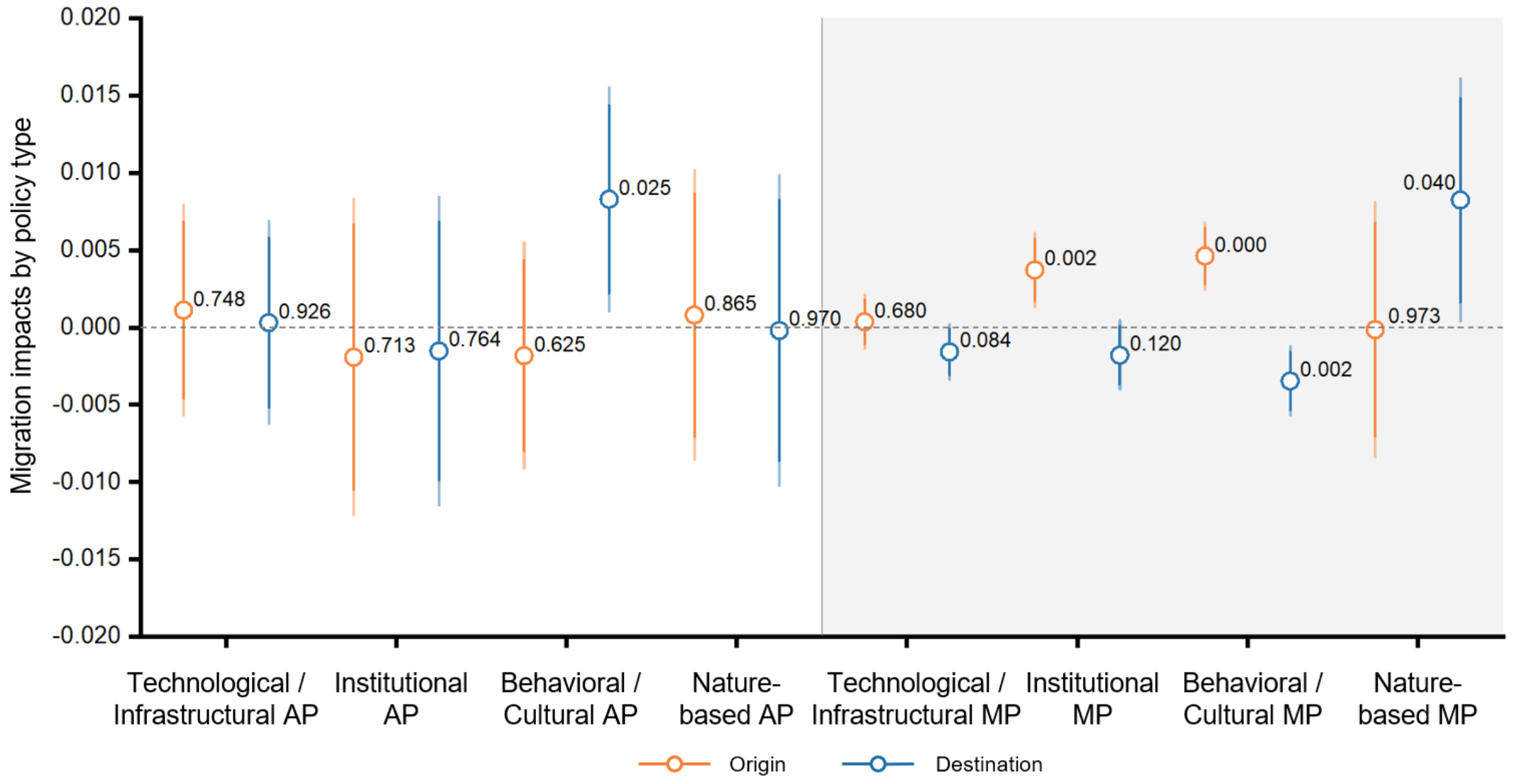

**Fig. 3 Heterogeneity in the effect of heat-related policies (HPs) on migration**. The $y$-axis shows the marginal effects and error bars of one additional specific type of policy on out-migration rates. The central estimates shown with circles are regression point estimates, neighboring numbers indicate significant $P$ values of estimates computed by t-tests, dark-colored lines indicate 90% confidence intervals, and light-colored lines indicate 95% confidence intervals. The yellow-colored effects represent the effect of HPs at the origin on the outflow rate of the origin, and the blue-colored effects represent the effect of HPs at the destination on the outflow rate of the origin. The transparent background on the left represents AP, and the grey background on the right represents MP.

**Table 1: Impacts of different types of heat-related policies on migration**

| | Outcome: ln (annual out-migration rate) |
|---|---|

*Corresponding author: Xing Su. Email address: 20250612@hhu.edu.cn

| | (1) | (2) | (3) |
|---|---|---|---|
| Technological/Infrastructural AP (origin) | 0.0011 | 0.0030 | 0.0068 |
| | (0.0035) | (0.0051) | (0.0123) |
| Institutional AP (origin) | -0.0019 | -0.0031 | -0.0107 |
| | (0.0052) | (0.0075) | (0.0151) |
| Behavioral/Cultural AP (origin) | -0.0018 | 0.0027 | 0.0014 |
| | (0.0037) | (0.0077) | (0.0165) |
| Nature-based AP (origin) | 0.0008 | -0.0011 | -0.0034 |
| | (0.0048) | (0.0085) | (0.0186) |
| Technological/Infrastructural MP (origin) | 0.0004 | -0.0044** | 0.0001 |
| | (0.0009) | (0.0018) | (0.0031) |
| Institutional MP (origin) | 0.0037*** | 0.0035 | 0.0010 |
| | (0.0012) | (0.0023) | (0.0060) |
| Behavioral/Cultural MP (origin) | 0.0046*** | 0.0044** | -0.0027 |
| | (0.0011) | (0.0017) | (0.0138) |
| Nature-based MP (origin) | -0.0001 | -0.0014 | 0.0046 |
| | (0.0042) | (0.0066) | (0.0149) |
| Technological/Infrastructural AP (destination) | 0.0003 | -0.0022 | 0.0069 |
| | (0.0033) | (0.0048) | (0.0129) |
| Institutional AP (destination) | -0.0015 | 0.0101 | 0.0054 |
| | (0.0051) | (0.0077) | (0.0145) |
| Behavioral/Cultural AP (destination) | 0.0083** | 0.0071 | 0.0102 |
| | (0.0037) | (0.0079) | (0.0146) |
| Nature-based AP (destination) | -0.0002 | 0.0051 | -0.0062 |
| | (0.0051) | (0.0083) | (0.0192) |
| Technological/Infrastructural MP (destination) | -0.0016* | 0.0047*** | 0.0062* |
| | (0.0009) | (0.0016) | (0.0032) |
| Institutional MP (destination) | -0.0018 | -0.0050** | -0.0042 |
| | (0.0012) | (0.0020) | (0.0059) |
| Behavioral/Cultural MP (destination) | -0.0035*** | -0.0053*** | -0.0068 |
| | (0.0011) | (0.0017) | (0.0136) |
| Nature-based MP (destination) | 0.0082** | 0.0073 | -0.0146 |
| | (0.0040) | (0.0066) | (0.0167) |
| F-test statistic | 92.3519 | 74.4978 | 78.7934 |
| F-test p-value | 0.0000 | 0.0000 | 0.0000 |
| Hausman test statistic | 34.8554 | 101.2359 | 71.4375 |
| Hausman test p-value | 0.0042 | 0.0000 | 0.0000 |
| Observations | 11,177 | 8,549 | 5,492 |
| Adjusted $R^2$ | 0.0227 | 0.0316 | 0.0319 |
| RMSE | 0.1778 | 0.1750 | 0.1427 |
| AIC | -13044.76 | -12474.50 | -14517.55 |
| BIC | -12927.62 | -12361.65 | -14411.78 |

Note: (1) main specification, (2) one-year lagged effects, (3) two-year lagged effects. Regression coefficients with

*Corresponding author: Xing Su. Email address: 20250612@hhu.edu.cn

standard errors in parentheses. We use cluster robust standard errors at the origin–destination pair level in all models. $P$ values of coefficients were derived based on two-sided t-tests and no adjustments were made for multiple comparisons. $^{*}P < 0.1$, $^{**}P < 0.05$ and $^{***}P < 0.01$. The F-test is conducted between the gravity-type fixed effects model and the corresponding pooled OLS model and the Hausmann test is conducted between the gravity-type fixed effects model and the corresponding random effects model. The adjusted $R^2$, root mean square error (RMSE), Akaike information criterion (AIC) and Bayesian information criterion (BIC) evaluate the performance and goodness of fit of the models.

## Non-linear contextual moderation of policy effects

Beyond the intrinsic differences among policy types, the realized effects of any policy are critically moderated by the contextual factors surrounding its implementation. Here, by examining the interactions between multiple HPs and median household income, aging rates, racial diversity, and educational attainment rates, we find significant variations in migration responses on the basis of these socioeconomic characteristics and the differences between them.

We begin with linear interaction models, which reveal that the moderating effects of income, racial diversity, and education are statistically significant, while the aging rate generally shows no significant moderating effects (**Supplementary Table S5**). This anomaly motivates us to test for non-linear interactions, acknowledging that the impact of context may not follow a strictly linear pattern. Therefore, we extend the model to include both linear and quadratic interaction terms. Model performance metrics—including adjusted $R^2$, RMSE, AIC, and BIC—consistently favor the quadratic interaction model (**Supplementary Tables S5 and S6**), supporting the presence of complex, non-linear moderating dynamics.

The marginal effects of origin-based HPs on out-migration vary significantly across socioeconomic dimensions (**Fig. 4a**). For ln(income), institutional MPs exhibit a U-shaped relationship: their marginal effects are significantly negative at both low (~11.0) and high (~11.7) income levels, and positive at middle-income levels (~11.4). This suggests that such policies may deter migration in economically polarized regions but drive migration in intermediate contexts. By contrast, behavioral/cultural MPs follow a monotonically declining trend, where the migration effect decreases with income and stabilizes at higher levels, indicating that these policies are less effective, or even counterproductive, in wealthier counties. Regarding aging rates, institutional MPs show an inverted U-shaped pattern, with a positive effect on migration at moderate aging levels (~0.16) but negative effects at lower values (~0.12). Conversely, behavioral/cultural MPs follow a U-shaped pattern, suggesting complex age-related dynamics. With respect to racial diversity, the migration effects of both institutional and behavioral/cultural MPs are insignificant at low diversity levels but become significantly positive as diversity increases (~0.5). This finding implies that MPs may be more migration-inducing in racially heterogeneous contexts, potentially reflecting differing social sensitivities or mobilization capacities. For educational attainment, institutional MPs again show a U-shaped effect, with significant positive impacts at both low (~0.30) and high (~0.60) education levels, and muted or negative impacts at mid-range values. This suggests that these policies may trigger out-migration in both under- and over-educated contexts, while stabilizing population retention in moderately educated regions. In contrast, behavioral/cultural MPs display a nearly flat to mildly upward-sloping curve, indicating limited or weak educational moderation.

At the destination, non-linear effects are generally less volatile for technological/infrastructural and behavioral/cultural MPs, while behavioral/cultural APs and nature-based MPs exhibit more pronounced variation (**Fig. 4b**). For example, in terms of ln(income), nature-based MPs show a strongly negative slope and the migration effect is significantly positive at lower income levels (~11.0). This suggests that low-income destinations implementing nature-based MPs are more attractive to migrants, while high-income destinations adopting the same measures may deter them. In terms of aging rates, nature-based MPs follow an inverted U-shaped pattern, producing positive migration effects at moderate aging levels (~0.16), and negative effects at both younger and older extremes. Other destination policies, such as behavioral/cultural APs, display flat or statistically insignificant trends across aging gradients. For educational attainment rates, behavioral/cultural APs show a monotonically increasing relationship with education, becoming significantly positive above ~0.5. Other policy types, however, remain relatively insensitive to variation in education, with marginal effects generally small and statistically insignificant.

*Corresponding author: Xing Su. Email address: 20250612@hhu.edu.cn

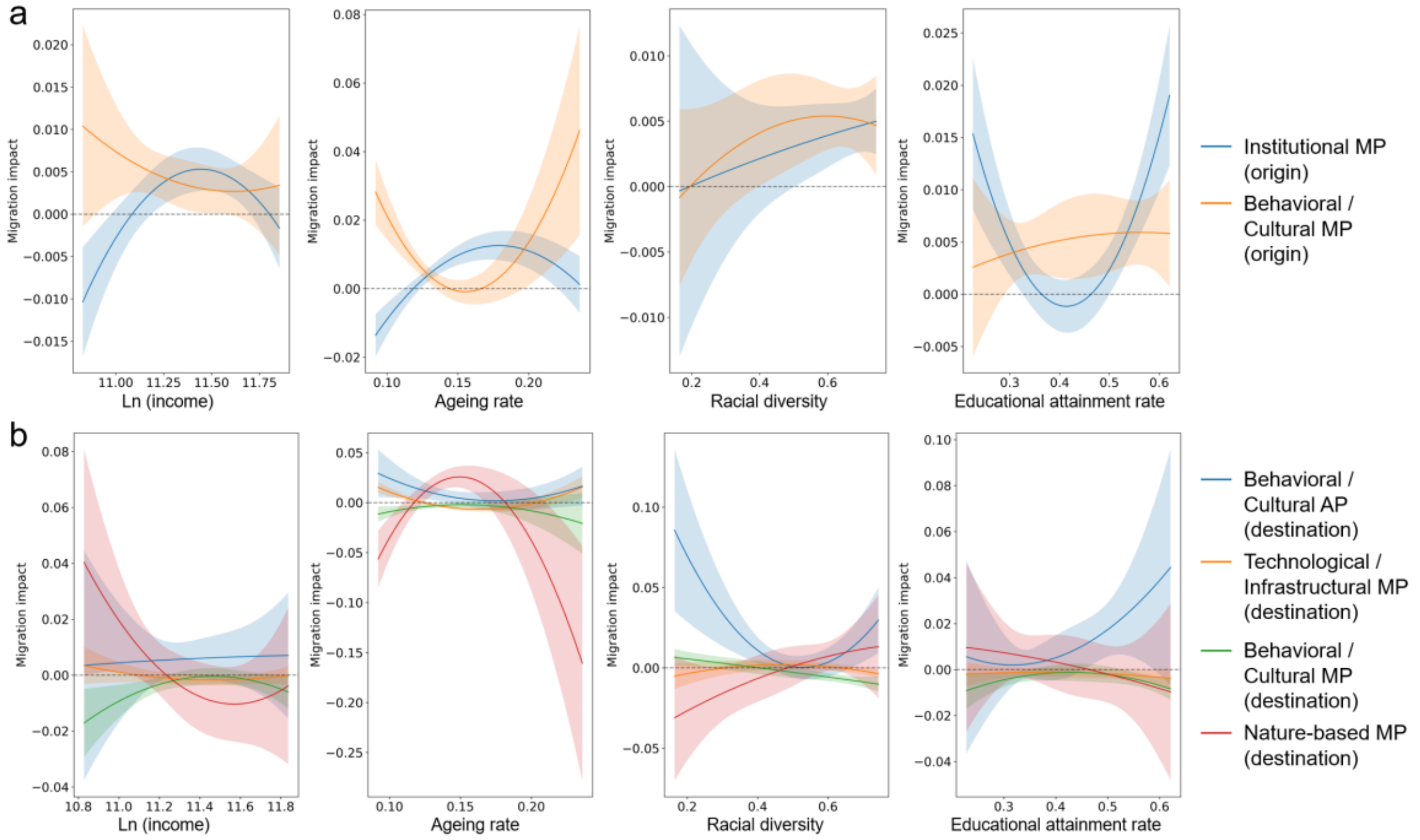

**Fig. 4 The role of localized characteristics in shaping migration responses to heat-related policies (HPs). a**, The role of origin characteristics in shaping migration responses to institutional MP and behavioral/cultural MP. **b**, The role of destination characteristics in shaping migration responses to behavioral/cultural AP, technological /infrastructural MP, behavioral/cultural MP and nature-based MP. The *y* axis shows the marginal effects of a 1 change in the specific type policy on the out-migration rate. These figures are estimated by using interaction models with quadratic terms based on interactions between HPs for the origin or destination and ln(income), the ageing rate, the racial diversity and the educational attainment rate (**Supplementary Table S6**). The increasing (decreasing) function indicates that the impact of migration increases (decreases) as the level of the interaction variable increases. We calculated the marginal effects using the interaction model when only the policy-migration relationship had significant coefficients, as shown in **Fig. 3**.

## Discussion

In this study, we estimate the causal impact of heat-related policies (HPs) on U.S. internal population migration using a series of gravity-type fixed-effects models. Our findings suggest that the implementation of heat mitigation policies (MPs) tends to drive out-migration, a result that is robust across multiple specifications. In contrast, heat adaptation policies (APs) are generally associated with in-migration, although their effects are less statistically significant. These divergent patterns invite consideration of potential underlying mechanisms.

*Corresponding author: Xing Su. Email address: 20250612@hhu.edu.cn

One plausible interpretation for the out-migration effect of MPs is their potential to generate short-term economic adjustments, in line with prior literature on the unintended consequences of environmental regulation[29–31]. Our heterogeneity analyses offer further indirect support to this perspective: Institutional and behavioral/cultural MPs—policy types most commonly associated with regulatory compliance costs or shifts in production/consumption patterns — exhibit the strongest positive association with out-migration in origin counties (**Table 1**). For example, policies like industrial emissions standards or carbon pricing are frequently linked to increased operational costs for businesses[32,33] and these extra costs may be partially passed through to households via higher prices or moderated wage growth, potentially resulting in economic disincentives to remain in the locality. This interpretation aligns with behavioral insights suggesting populations may exhibit heightened sensitivity to immediate economic pressures relative to longer-term, albeit significant, climate risks[34]. Interestingly, we also find that nature-based MPs implemented in destination counties are positively associated with in-migration from origin counties. This contrasting effect suggests that the nature of the MP matters. Unlike institutional and behavioral/cultural MPs, these policies may not impose immediate financial burdens, but instead improve environmental quality and local livability—functionally resembling APs in their short-term effects.

We also investigate how local socio-economic contexts moderate the effects of HPs on migration. These moderating effects are both pronounced and non-linear, especially regarding aging rates. For example, at origin counties, institutional MPs exert a significantly positive effect on migration at moderate ageing levels (~0.16), whereas effects become negative at low (~0.12) ends. This reflects the situation in which the older population exhibits a stronger out-migration response to the institutional MPs. This finding is similar to the findings of Hoffmann et al.[11], who observed that older individuals tend to exhibit stronger migration responses to drought and aridity.

Our study contributes to the emerging literature on climate policy and migration in several important ways. First, while much of the existing research focuses on the climate–migration nexus, few studies isolate the impact of climate policy itself. We address this gap by integrating policy data from CDP with IRS migration data in a gravity-type model framework. Second, we examine heterogeneous effects across HP types using machine learning–enhanced classification. Third, we identify how local socio-economic characteristics—including income, education, aging, and racial diversity—moderate HP effects on migration. Such heterogeneous impacts that stem from policy types and the moderating effects of local characteristics can inform more nuanced and targeted policy decisions that balance environmental goals with potential demographic consequences.

Our study also presents several limitations. First, data constraints may introduce selection bias. Although the CDP dataset is among the most comprehensive sources of climate policy data, it cannot guarantee that all implemented HPs are recorded, nor does it uniformly detail their implementation stages or precise geographic scopes within a county. To mitigate this, we merged policy data in county scale and restrict the analysis to origin-destination pairs where both counties report non-zero HPs. While this enhances comparability, it may reduce the generalizability of our findings to areas with little or no policy activity. Second, our empirical design estimates average treatment effects, but does not disentangle the direct and indirect pathways through which HPs influence migration. Further work incorporating finer-grained data (e.g., firm-level, household surveys) and mediation analysis techniques could help unpack these mechanisms. Third, the scarcity of directly comparable empirical studies makes it difficult to benchmark our effect sizes against an established consensus. This is because the growing body of related work, such as the structural simulation models[35,36], pursues a different objective: projecting migration under long-term, hypothetical policy scenarios rather than estimating the causal effect of policies already implemented. Consequently, while our study provides among the first empirical estimates of these effects, the direct comparability of our results with the existing modeling literature is limited due to fundamental differences in methodology and purpose.

*Corresponding author: Xing Su. Email address: 20250612@hhu.edu.cn

While there are still many unanswered questions about the nexus between climate change, climate policy, and migration, this study highlights the pivotal role of climate policy in influencing migration. To date, empirical research on this topic remains scarce. Moreover, the fact that climate change and climate policy will continue co-prosperity worldwide for the foreseeable future emphasizes the urgency of studying the heterogeneous migration effects of diverse climate policies in various human and geographic contexts. Therefore, future research should investigate the spatiotemporally refined, pathway-diverse causal effects of climate policies on migration in different countries and regions under multiple climate hazard contexts.

## Materials and methods

### Data

The climate policy data used in this study are derived from the CDP cities' mitigation and adaptation actions dataset[37] (formerly the Carbon Disclosure Project). CDP maintains an extensive repository of climate-related datasets, publicly accessible through its website (available at https://data.cdp.net/). The CDP data are collected via the CDP-ICLEI Unified Reporting System and include actions disclosed by cities or counties to CDP or one or more global forums, such as Local Governments for Sustainability, the Global Covenant of Mayors for Climate & Energy, or C40 Cities[38]. Specifically, we used the 2017-2020 mitigation and adaptation actions dataset, which includes information such as the reporting time and place, place coordinates, the year the actions were undertaken, and the names and descriptions of the actions. We first filtered out the climate actions implemented in the U.S. and then identified those specifically related to heat. Since MPs are aimed at reducing $CO_2$ emissions to slow the pace of warming, we included all mitigation actions into our analyses ($n = 3{,}945$). Heat adaptation policies primarily focus on interventions that alleviate heat stress, so we used keywords such as "heat," "high temperature," and "drought" to filter APs ($n = 768$). We integrated the CDP data to the corresponding counties using the Geopandas library in Python. The geodatabases used were the 2020 TIGER/Line shapefiles for US geographical boundaries at the county level, provided by the US Census Bureau (available at https://www.census.gov/geographies/ mapping-files/ time-series/geo/carto-boundary-file.html). We merged the columns for "action" and "action description" to obtain comprehensive descriptive text for each action to support subsequent classification.

The migration data used in this study were derived from the 2017-2021 IRS county-to-county migration estimates[39]. IRS migration data (available at https://www.irs.gov/statistics/soi-tax-stats-migration-data) have been widely utilized in the literature and report the total number of movers entering and leaving each county on an annual basis. IRS data are based on year-to-year address changes reported on individual income tax returns, and it is the most complete and largest dataset on migration within the U.S.[40] There are two main alternatives available for migration data[24]: survey data from the U.S. Census Bureau that measure migration (e.g., the American Community Survey (ACS) or the Current Population Survey), which provide longer time scales, such as the ACS data[41], providing county-to-county migration data every five years, which does not coincide with the time period of this study; data from the Federal Reserve Bank of New York/Equifax Consumer Credit Panel (CCP) migration data, the CCP data represents only those U.S. adults with Social Security numbers (SSNs) and credit histories and relies on consumer credit histories to infer likely addresses[24]. As a result, the dataset suffers from insufficient data coverage and incorrect selection of migratory addresses. Taking into account the above trade-offs, the IRS migration data is the most appropriate dataset to use for this study. Moreover, this study only considers the contiguous US, which includes the lower 48 states and Washington DC. Other territories are excluded because they are outliers in terms of geolocations, demographic factors, and migration patterns.

*Corresponding author: Xing Su. Email address: 20250612@hhu.edu.cn

The demographic data we used came from the 5-year estimates 2017-2020 ACS (available at https://www.census. gov/). In line with the literature on climate-related migration[42], we selected income as the economic background, and aging rate, educational attainment rate as social characteristics. Additionally, we accounted for the role of social connectivity in influencing migration decisions by including racial diversity as an indicator, with the specific calculation method based on ref.[41].

We have compiled the essential information on key variables, including their definitions, measurement units and sources, in **Extended Data Table 1**. The data were processed by utilizing Python version 3.9.7.

**Machine-learning classifiers for policy types**

In this study, we fine-tune a pre-trained DistilBERT model using our annotated dataset and apply the model with the best performance to predict the categories of all remaining HPs. We first draw upon the classification framework for climate adaptation actions proposed by Lea Berrang-Ford et al.[7], categorizing HPs, both APs and MPs, into technological/infrastructural, institutional, behavioral/cultural, and nature-based categories. Detailed definitions and examples for each category are provided in the **Supplementary Note S1**. To train the classifier more efficiently and enhance the generalization ability of the classifier, we first identify the duplicates after screening 4,713 HPs, and then augment them using synonym replacement and sentence rearrangement. We then randomly select 20% of the 4,713 HPs for manual annotation to determine their respective types. Notably, a single climate action could belong to multiple categories; for instance, some cities implement both nature-based solutions and public education as part of their heat adaptation strategies. Our annotation team consists of three members. The randomly selected samples are first independently reviewed by each team member, where they assign a value of 1 if the climate action belongs to a particular category and 0 otherwise. Subsequently, team members discuss cases with differing labels until a consensus is reached to reduce bias and ensure consistency across annotators. This process results in a training dataset for the machine learning classifiers.

We then train the classifier based on the DistilBERT model, which retains 97% of BERT's language understanding capabilities while significantly reducing computational resource requirements, a feature that has already proven successful in climate research[27]. We do not select ChatGPT or other large language models like BERT for two primary reasons: 1) large language models can have non-trivial climate impacts[27], and 2) balancing the task size with computational resources. We set the model's prediction threshold at 0.5, such that if the predicted probability of a climate action belonging to a specific category exceeded 0.5, it is assigned a value of 1, otherwise, it is assigned 0. Given this threshold, our classifier can automatically filter out HPs with semantically unclear or extremely short descriptions and classify them as not belonging to any type of the four categories. The model's performance metrics are then evaluated by comparing the predicted classifications with the annotated data.

*Corresponding author: Xing Su. Email address: 20250612@hhu.edu.cn

The training process for our model is informed by the methods outlined by Max Callaghan et al.[27] We conduct five-fold internal cross-validation for each hyperparameter combination to ensure generalizability, and the mean metric values of model performance from the five folds are used for external comparison. As Max Callaghan et al. noted, although grid search for hyperparameter optimization is transparent, robust, and exhaustive, it is computationally expensive, and alternative approaches like random search may offer some improvements. Therefore, to achieve optimal training and classification results, we evaluate three common hyperparameter optimization strategies: grid search, random optimization, and Bayesian optimization.

The results show that classifiers optimized via grid search outperform those optimized via the other two methods across all evaluation metrics, achieving a mean five-fold F1 score of 0.9274 and a mean ROC AUC score of 0.9771 (**Extended Data Fig. 2**). The best learning rate found using the grid search method is 0.0001, the optimal number of training epochs is 5, and the best batch size is 8. Finally, based on the best-performing classifier, we predict the categories for all remaining HPs.

**Main specification of statistical model**

We first identify counties that have implemented HPs based on CDP data, then align these counties to the IRS database for county-pair migration as origin and destination, respectively, and organize them uniformly in the perspective of out-migration. To prevent missing data due to the lack of reporting of HPs to the CDP by some counties, we use only pairs of counties in which none of the HPs are zero in our analyses.

To estimate the impacts of HPs on the bilateral out-migration rate between regions of origin $i$ and destination $j$, we use a series of gravity-type fixed-effects models[43]. Our models are estimated parsimoniously to avoid overcontrolling issues[11,44]. The full model is shown as:

$$\ln(y_{ijt}) = \partial_0 + \sum_c \beta_c hp_{c,it} + \sum_c \gamma_c hp_{c,jt} + u_{ij} + v_t + \varepsilon_{ijt} \quad (1)$$

Where $y_{ijt}$ indicates the annual out-migration rate from the origin county $i$ to the destination county $j$ at time $t$. $c$ indicates category of the HP, in this study, i.e., adaptation policies (APs) and mitigation policies (MPs). $hp_{c,it}$ and $hp_{c,jt}$ refer to the amounts of APs and MPs in the origin and destination, respectively. **Supplementary Table S2** provides summary statistics of above variables. $u_{ij}$ is a set of dummy variables, referring to the origin and destination pair fixed effects. These spatial 'fixed effects' account for relatively stable characteristics of county pairs over time, such as the political contexts, climatic characteristics and cultural differences. $v_t$ indicates a set of year-of-sample 'fixed effects', which are a set of dummy variables for each year in the 2017–2021 sample. This time fixed effects account for the confounding influences of temporal trends. The term $\varepsilon_{ijt}$ indicates model error terms. $\partial_0$ indicates the intercept of the linear parts. To estimate robust standard errors, all standard errors were clustered at the origin–destination pair level.

*Corresponding author: Xing Su. Email address: 20250612@hhu.edu.cn

This study establishes the validity of our model in identifying causal effects through rigorous econometric tests. First, we confirm that the core explanatory variables (the amount of AP and MP at origin and destination) are sufficiently dynamic through within-group variation analysis: panel data strength assessment shows that 69.7% of county pairs have observations of 2 years and more, and the proportion of within-group variation is greater than 30% for all core variables (AP at origin = 45.2%, MP at origin = 49.5%, AP at destination = 45.1%, MP at destination = 51.0%), which meets the fixed-effects model's requirements for time variability. Second, the F-test results (F = 93.2875, $P$ = 0.000) significantly rejected the pooled OLS model at the 1% level, confirming the necessity of county pair fixed effects. The Hausman test ($\chi^2$ = 9.0866, $P$ = 0.059) significantly rejected the random effects hypothesis at the 10% significance level, indicating that the individual effects are correlated with the explanatory variables, at which point the fixed effects model provides consistent estimates. Therefore, the gravity-type fixed-effects model we chose satisfies the conditional independence assumption and its coefficients can be explained as the causal effect of the core variables on migration flows.

**Heterogeneous effects by policy types**

To delve into the precise impacts of different HPs on migration, we estimate the heterogeneity of the effect across policy types using the following formula:

$$\ln(y_{ijt}) = \beta_0 + \sum_c \sum_e \emptyset_{ce} \times hp_{ce,it} + \sum_c \sum_e \varphi_{ce} \times hp_{ce,jt} + u_{ij} + v_t + \varepsilon_{ijt} \quad (2)$$

where $e$ indicates a specific type of heat-related policies. There are four policy types, namely technological/infrastructural, institutional, behavioral/cultural and nature-based. Both AP and MP are categorized in the above four types. $\emptyset_{ce}$ denotes the type-specific effect for the category $c$ of heat-related policy. $hp_{ce,it}$ and $hp_{ce,jt}$ refer to the amounts of the specific policy type of APs and MPs in the origin and destination, respectively. Other parts of the model remain the same as in equation (1).

**Moderating effects by contextual factors**

We use interaction models to investigate how socio-economic characteristics of origin and destination counties shape the migration response. To simplify the calculations, we hereby consider in this step only the significant $\emptyset_{ce}$ and $\varphi_{ce}$ estimated by equation (2).

We first take linear interactions into account; the whole model can be written as:

$$\ln(y_{ijt}) = \beta_0 + \sum_c \sum_e \emptyset_{ce} \times hp_{ce,it} + \sum_c \sum_e \varphi_{ce} \times hp_{ce,jt} + \beta_1 \times M_i + \beta_2 \times M_j + \sum_c \sum_e \theta_{ce} \times (M_i \times hp_{ce,it}) + \sum_c \sum_e \vartheta_{ce} \times (M_j \times hp_{ce,jt}) + u_{ij} + v_t + \varepsilon_{ijt} \quad (3)$$

where $M_i$ is the socio-economic indicators of origin county, $M_j$ is the socio-economic indicators of destination county. We choose ln(income), ageing rate, educational attainment rate and racial diversity as the four socio-economic indicators. For each indicator, we compute the coefficient results using the interaction model, which leads to the calculation of conditional marginal effects.

In addition, we consider more complex quadratic interactions, i.e., we incorporate both linear interaction terms and quadratic interaction terms into the interaction model. **Supplementary tables S5 and S6** provide the regression results of linear interaction models and non-linear interaction models, respectively.

*Corresponding author: Xing Su. Email address: 20250612@hhu.edu.cn

## Data availability

All the data utilized in this study are based on publicly available data sources as described in the Methods and are available from the corresponding references.

*Corresponding author: Xing Su. Email address: 20250612@hhu.edu.cn

*Corresponding author: Xing Su. Email address: 20250612@hhu.edu.cn

*Corresponding author: Xing Su. Email address: 20250612@hhu.edu.cn

*Corresponding author: Xing Su. Email address: 20250612@hhu.edu.cn

*Corresponding author: Xing Su. Email address: 20250612@hhu.edu.cn

*Corresponding author: Xing Su. Email address: 20250612@hhu.edu.cn

## Extended data figures and tables

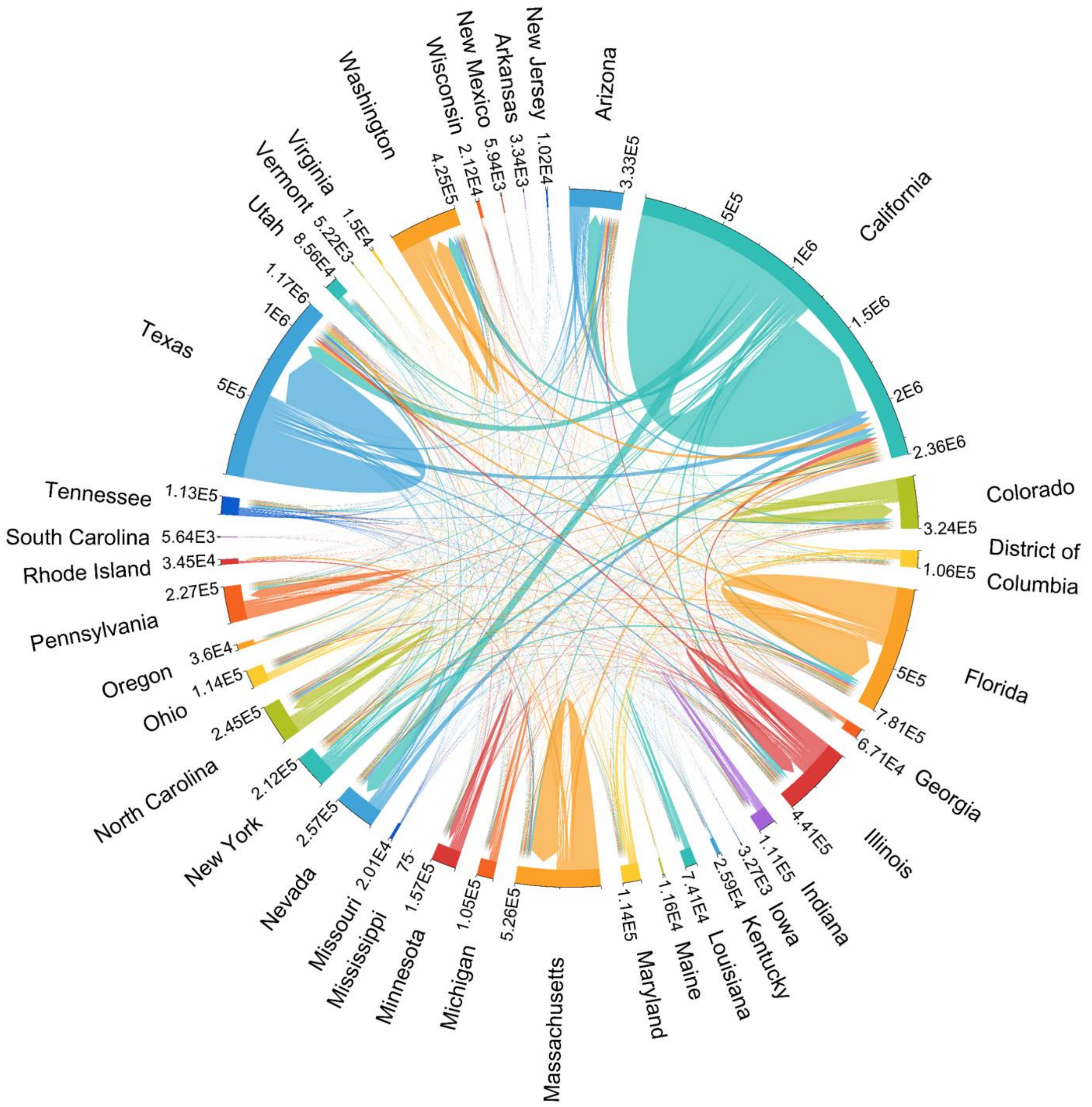

**Extended Data Fig. 1 Migration flows between U.S. states with heat-related policies (HPs) implemented in 2017-2020.**

*Corresponding author: Xing Su. Email address: 20250612@hhu.edu.cn

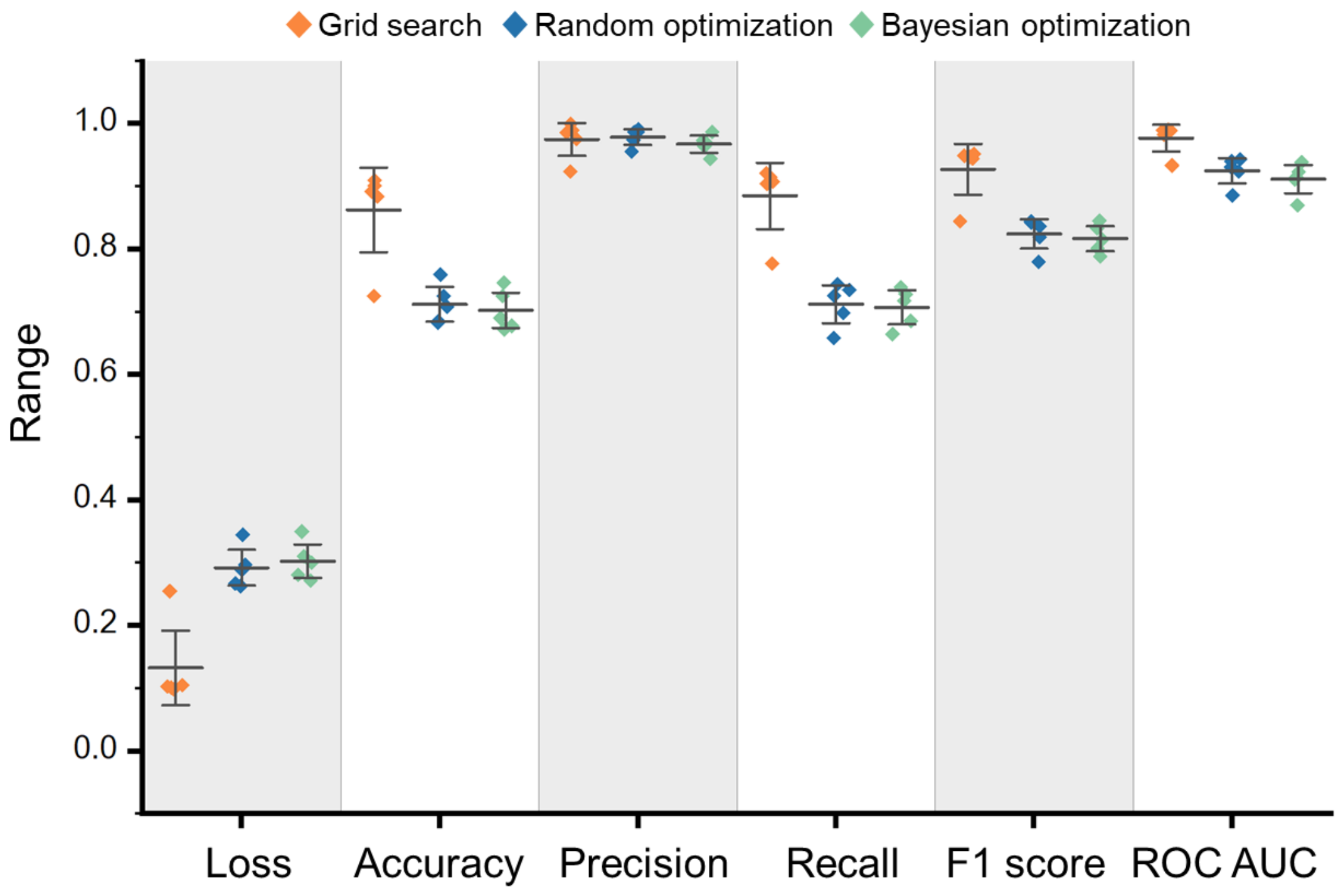

**Extended Data Fig. 2 DistilBERT model performance.** Loss is used to estimate the degree of inconsistency between the model's predicted values and the true values. Accuracy indicates the proportion of samples that are correctly classified. Precision is the proportion of samples that truly belong to the positive class out of all samples predicted to be in the positive class. Recall is defined as the proportion of samples correctly predicted to be in the positive class out of all samples in the positive class. Recall is defined as the proportion of samples correctly predicted to be in the positive class out of all samples in the positive class. F1 score is the reconciled average of Precision and Recall, which is used to evaluate the performance of the model in a comprehensive way. ROC AUC (Area Under the Curve) is the area under the curve of ROC (Receiver Operating Characteristic), which is used to quantify the overall performance of the classifiers. The three horizontal lines from the top to the bottom of each box represent the upper 95% confidence interval, the mean and the lower 95% confidence interval, respectively.

*Corresponding author: Xing Su. Email address: 20250612@hhu.edu.cn

**Extended Data Table 1 Data variables and sources.**

| Variable | Definition | Unit of measurement | Data source |
|---|---|---|---|
| **Socio-economic variables** | | | |
| Income | Median household income | ln [Median income (US$ per household)] | American Community Survey (ACS) |
| Racial diversity | The probability of two randomly selected individuals from a specific location belonging to different races. | $RD_i = 1 - \sum_i p_i^2$, where $p_i$ represents the proportion of each racial group within county $i$. | American Community Survey (ACS) |
| Ageing rate | Percent population 65 years and over | Percentage | American Community Survey (ACS) |
| Educational attainment rate | Percent bachelor's degree or higher of population 25 years and over | Percentage | American Community Survey (ACS) |
| **Policy variables** | | | |
| AP | Number of heat adaptation policies | Number | Carbon Disclosure Project (CDP) |
| MP | Number of heat mitigation policies | Number | Carbon Disclosure Project (CDP) |
| HP | Number of heat-related policies | Number | Carbon Disclosure Project (CDP) |
| **Migration variables** | | | |
| Outflow rate | Outflow rate of from origin to destination | ln [Outflow rate] | Internal Revenue Service (IRS) |

*Corresponding author: Xing Su. Email address: 20250612@hhu.edu.cn

**Supplementary Information for**

Unveiling contrasting impacts of heat mitigation and adaptation policies on U.S. internal migration

This file includes:
Supplementary Note S1
Tables S1 to S6
Figures S1 to S5
References

*Corresponding author: Xing Su. Email address: 20250612@hhu.edu.cn

## Contents

*Corresponding author: Xing Su. Email address: 20250612@hhu.edu.cn

## 1.Supplementary Notes

**Note S1** Definitions and examples for four types of heat-related policies

Lea Berrang-Ford et al.[1] used supervised machine learning to screen the literature published between 2013 and 2019 to systematically inventory global evidence of human adaptation to climate change. They classified climate adaptation actions into four categories: technological/infrastructural, institutional, behavioral/cultural, and nature-based. This study follows this categorization for heat-related policies (HP) and extends it to include both adaptation and mitigation policies. Heat mitigation policies (MP) are centered around carbon dioxide ($CO_2$) emission reduction as a means of slowing down the rate of global warming[2], while heat adaptation policies (AP) take the form of interventions to reduce people's heat stress[3]. Detailed definitions and relevant examples of the four subtypes for both AP and MP are as follows.

Technological/infrastructural policies involve the implementation of solutions that utilize technology and physical infrastructure to reduce the frequency, intensity and adverse impact of extreme heat events and heatwaves. For mitigation, an example could involve the deployment of solar panels and wind power systems to generate clean energy to reduce carbon emission. In terms of adaptation, infrastructure upgrades like developing heat-resilient building materials and constructing cooling centers (grey infrastructure) to provide cooling effects in cities.

Institutional policies are governance frameworks and regulatory measures aimed at addressing heat-related risks and vulnerabilities within societies. These policies establish guidelines for heat emergency response plans, heat-health action strategies, and coordination mechanisms between governmental agencies, healthcare providers, and community organizations. In the context of mitigation, establishing building codes that mandate energy-efficient cooling systems can significantly reduce heat-related energy consumption. Regarding adaptation, instituting heat action plans to protect vulnerable populations during extreme heat events illustrates an institutional approach to enhancing resilience.

Behavioral/cultural policies focus on changing individual and societal behaviors and norms to reduce $CO_2$ emissions and heat exposure during periods of extreme heat. These policies aim to raise awareness about heat-related risks, educate communities on heat safety measures, and encourage behaviors that are benefit for heat mitigation and adaptation. For mitigation, promoting public awareness campaigns encouraging energy conservation can help reduce electricity demand. In terms of adaptation, fostering community programs that educate individuals on heat-related health risks and methods to stay cool during heatwaves can enhance preparedness and resilience.

Nature-based policies leverage natural ecosystems and green infrastructure to mitigate $CO_2$ emissions and the impacts of extreme heat events. These policies focus on promoting carbon sinks and biodiversity, restoring natural habitats, and providing shade, cool urban areas to improve thermal comfort. In terms of mitigation, the policies on planting more trees in urban areas to absorb $CO_2$ and enhance carbon sinks. For adaptation, restoring and preserving green spaces such as parks and urban forests can help mitigate the adverse effects of heatwaves by providing natural cooling.

It is notable that the types of climate policies are not exclusive, i.e., a single climate policy may encompass multiple types. For example, in October 2019, San José Clean Energy (SJCE) secured City Council approval to participate in the state's California Electric Vehicle Infrastructure Project

*Corresponding author: Xing Su. Email address: 20250612@hhu.edu.cn

(CALeVIP)[4]. This project involves not only the construction of charging infrastructure but also a vehicle charger incentive program to mitigate $CO_2$ emissions.

*Corresponding author: Xing Su. Email address: 20250612@hhu.edu.cn

## 2.Supplementary tables and figures

### 2.1. Tables S1 to S6

**Table S1 List of 106 origin counties**

| FIPS code | County name | Located state | Located region | FIPS code | County name | Located state | Located region |
|---|---|---|---|---|---|---|---|
| 4005 | Coconino County | Arizona | West | 28001 | Adams County | Mississippi | South |
| 4013 | Maricopa County | Arizona | West | 29019 | Boone County | Missouri | Midwest |
| 5143 | Washington County | Arkansas | South | 29510 | St. Louis city | Missouri | Midwest |
| 6001 | Alameda County | California | West | 32003 | Clark County | Nevada | West |
| 6013 | Contra Costa County | California | West | 32031 | Washoe County | Nevada | West |
| 6019 | Fresno County | California | West | 34013 | Essex County | New Jersey | Northeast |
| 6029 | Kern County | California | West | 35049 | Santa Fe County | New Mexico | West |
| 6037 | Los Angeles County | California | West | 36001 | Albany County | New York | Northeast |
| 6059 | Orange County | California | West | 36029 | Erie County | New York | Northeast |
| 6067 | Sacramento County | California | West | 36061 | New York County | New York | Northeast |
| 6073 | San Diego County | California | West | 37021 | Buncombe County | North Carolina | South |
| 6081 | San Mateo County | California | West | 37063 | Durham County | North Carolina | South |
| 6085 | Santa Clara County | California | West | 37067 | Forsyth County | North Carolina | South |
| 6087 | Santa Cruz County | California | West | 37081 | Guilford County | North Carolina | South |
| 6095 | Solano County | California | West | 37119 | Mecklenburg County | North Carolina | South |
| 6113 | Yolo County | California | West | 37135 | Orange County | North Carolina | South |
| 8013 | Boulder County | Colorado | West | 37183 | Wake County | North Carolina | South |
| 8031 | Denver County | Colorado | West | 39049 | Franklin County | Ohio | Midwest |
| 8037 | Eagle County | Colorado | West | 39061 | Hamilton County | Ohio | Midwest |

*Corresponding author: Xing Su. Email address: 20250612@hhu.edu.cn

| | | | | | | | |
|---|---|---|---|---|---|---|---|
| 8059 | Jefferson County | Colorado | West | 41029 | Jackson County | Oregon | West |
| 8097 | Pitkin County | Colorado | West | 41039 | Lane County | Oregon | West |
| 11001 | District of Columbia | District of Columbia | South | 41047 | Marion County | Oregon | West |
| 12011 | Broward County | Florida | South | 42003 | Allegheny County | Pennsylvania | Northeast |
| 12086 | Miami-Dade County | Florida | South | 42071 | Lancaster County | Pennsylvania | Northeast |
| 12095 | Orange County | Florida | South | 42091 | Montgomery County | Pennsylvania | Northeast |
| 12099 | Palm Beach County | Florida | South | 42095 | Northampton County | Pennsylvania | Northeast |
| 13051 | Chatham County | Georgia | South | 42101 | Philadelphia County | Pennsylvania | Northeast |
| 13121 | Fulton County | Georgia | South | 44007 | Providence County | Rhode Island | Northeast |
| 17031 | Cook County | Illinois | Midwest | 45079 | Richland County | South Carolina | South |
| 17119 | Madison County | Illinois | Midwest | 47037 | Davidson County | Tennessee | South |
| 18057 | Hamilton County | Indiana | Midwest | 47093 | Knox County | Tennessee | South |
| 18097 | Marion County | Indiana | Midwest | 47157 | Shelby County | Tennessee | South |
| 18105 | Monroe County | Indiana | Midwest | 48029 | Bexar County | Texas | South |
| 19101 | Jefferson County | Iowa | Midwest | 48061 | Cameron County | Texas | South |
| 19103 | Johnson County | Iowa | Midwest | 48113 | Dallas County | Texas | South |
| 21111 | Jefferson County | Kentucky | South | 48121 | Denton County | Texas | South |
| 22051 | Jefferson Parish | Louisiana | South | 48141 | El Paso County | Texas | South |
| 22071 | Orleans Parish | Louisiana | South | 48201 | Harris County | Texas | South |
| 23005 | Cumberland County | Maine | Northeast | 48439 | Tarrant County | Texas | South |
| 24031 | Montgomery County | Maryland | South | 48453 | Travis County | Texas | South |
| 24510 | Baltimore city | Maryland | South | 49019 | Grand County | Utah | West |
| 25005 | Bristol County | Massachusetts | Northeast | 49035 | Salt Lake County | Utah | West |
| 25009 | Essex County | Massachusetts | Northeast | 49043 | Summit County | Utah | West |

*Corresponding author: Xing Su. Email address: 20250612@hhu.edu.cn

| | | | | | | | |
|---|---|---|---|---|---|---|---|
| 25015 | Hampshire County | Massachusetts | Northeast | 50007 | Chittenden County | Vermont | Northeast |
| 25017 | Middlesex County | Massachusetts | Northeast | 51121 | Montgomery County | Virginia | South |
| 25025 | Suffolk County | Massachusetts | Northeast | 51540 | Charlottesville County | Virginia | South |
| 26161 | Washtenaw County | Michigan | Midwest | 51760 | Richmond County | Virginia | South |
| 26163 | Wayne County | Michigan | Midwest | 51770 | Roanoke County | Virginia | South |
| 27053 | Hennepin County | Minnesota | Midwest | 53033 | King County | Washington | West |
| 27059 | Isanti County | Minnesota | Midwest | 53053 | Pierce County | Washington | West |
| 27123 | Ramsey County | Minnesota | Midwest | 53073 | Whatcom County | Washington | West |
| 27169 | Winona County | Minnesota | Midwest | 55025 | Dane County | Wisconsin | Midwest |

*Corresponding author: Xing Su. Email address: 20250612@hhu.edu.cn

**Table S2 Summary statistics of different heat-related policies**

| Policy type | Mean | SD | Min | Max |
|---|---|---|---|---|
| Technological/Infrastructural AP (origin) | 1.30 | 1.80 | 0 | 11 |
| Institutional AP (origin) | 0.51 | 0.91 | 0 | 5 |
| Behavioral/Cultural AP (origin) | 0.76 | 1.22 | 0 | 8 |
| Nature-based AP (origin) | 0.36 | 0.75 | 0 | 4 |
| Total AP (origin) | 2.92 | 4.01 | 0 | 22 |
| Technological/Infrastructural MP (origin) | 4.67 | 6.80 | 0 | 57 |
| Institutional MP (origin) | 2.45 | 4.02 | 0 | 30 |
| Behavioral/Cultural MP (origin) | 1.54 | 2.90 | 0 | 21 |
| Nature-based MP (origin) | 0.44 | 0.90 | 0 | 8 |
| Total MP (origin) | 9.10 | 12.91 | 0 | 84 |
| Total HP (origin) | 12.02 | 14.52 | 1 | 93 |
| Technological/Infrastructural AP (destination) | 1.79 | 0.90 | 0 | 6.4 |
| Institutional AP (destination) | 0.71 | 0.36 | 0 | 2.8 |
| Behavioral/Cultural AP (destination) | 1.07 | 0.49 | 0 | 3.2 |
| Nature-based AP (destination) | 0.53 | 0.26 | 0 | 2 |
| Total AP (destination) | 4.10 | 1.92 | 0 | 13.6 |
| Technological/Infrastructural MP (destination) | 5.64 | 3.20 | 0 | 16.2 |
| Institutional MP (destination) | 3.17 | 1.83 | 0 | 10 |
| Behavioral/Cultural MP (destination) | 1.93 | 1.58 | 0 | 8.3 |
| Nature-based MP (destination) | 0.52 | 0.35 | 0 | 2.4 |
| Total MP (destination) | 11.26 | 5.85 | 0 | 28.2 |
| Total HP (destination) | 15.36 | 6.49 | 2 | 41.8 |

*Corresponding author: Xing Su. Email address: 20250612@hhu.edu.cn

**Table S3 Summary statistics of variables**

| | Mean | SD | Min | Max |
|---|---|---|---|---|
| **Contextual variables** | | | | |
| Ln (Median income) (origin) | 11.296 | 0.231 | 10.464 | 11.905 |
| Ageing rate (origin) | 0.141 | 0.028 | 0.089 | 0.243 |
| Racial diversity (origin) | 0.444 | 0.166 | 0.086 | 0.747 |
| Educational attainment rate (origin) | 0.395 | 0.106 | 0.158 | 0.630 |
| Ln (income) (destination) | 11.326 | 0.071 | 11.054 | 11.567 |
| Ageing rate (destination) | 0.137 | 0.008 | 0.098 | 0.162 |
| Racial diversity (destination) | 0.505 | 0.043 | 0.303 | 0.652 |
| Educational attainment rate (destination) | 0.405 | 0.028 | 0.303 | 0.527 |
| **Outcome variable** | | | | |
| Outflow rate | 0.013 | 0.008 | 0.002 | 0.044 |

*Corresponding author: Xing Su. Email address: 20250612@hhu.edu.cn

**Table S4 Comparison results of estimated impacts on out-migration rate for one additional heat-related policy (HP)**

| | **Outcome: ln (annual out-migration rate)** | | | | | | | |
|---|---|---|---|---|---|---|---|---|
| | **(1)** | **(2)** | **(3)** | **(4)** | **(5)** | **(6)** | **(7)** | **(8)** |
| Total AP (origin) | -0.0002 | -0.0018** | -0.0004 | -0.0010 | -0.0002 | -0.0004 | 0.0019 | -0.0001 |
| | (0.0010) | (0.0007) | (0.0010) | (0.0010) | (0.0012) | (0.0014) | (0.0013) | (0.0028) |
| Total MP (origin) | 0.0021*** | 0.0020*** | 0.0020*** | 0.0019*** | 0.0020*** | 0.0025*** | 0.0003 | 0.0002 |
| | (0.0004) | (0.0002) | (0.0003) | (0.0004) | (0.0004) | (0.0005) | (0.0004) | (0.0010) |
| Total AP (destination) | 0.0020** | 0.0005 | 0.0020** | 0.0017 | 0.0025* | 0.0003 | 0.0028* | 0.0047* |
| | (0.0010) | (0.0007) | (0.0010) | (0.0011) | (0.0013) | (0.0014) | (0.0016) | (0.0028) |
| Total MP (destination) | -0.0015*** | -0.0015*** | -0.0015*** | -0.0014*** | -0.0016*** | -0.0014*** | -0.0007 | -0.0002 |
| | (0.0003) | (0.0002) | (0.0003) | (0.0003) | (0.0004) | (0.0004) | (0.0004) | (0.0011) |
| F-test statistic | 93.2875 | | 89.9530 | 48.6167 | 81.1055 | 77.6038 | 73.7065 | 79.4772 |
| F-test p-value | 0.0000 | | 0.0000 | 0.0000 | 0.0000 | 0.0000 | 0.0000 | 0.0000 |
| Hausman test statistic | 9.0866 | | 8.7486 | 9.7427 | 10.6773 | 11.2529 | 19.7966 | 36.3194 |
| Hausman test p-value | 0.0590 | | 0.0677 | 0.0450 | 0.0304 | 0.0239 | 0.0005 | 0.0000 |
| Origin and destination FE | Yes | | Yes | Yes | Yes | Yes | Yes | Yes |
| Year FE | Yes | | Yes | Yes | Yes | Yes | Yes | Yes |
| Observations | 11,177 | 11,177 | 11,177 | 10,060 | 8,942 | 7,924 | 8,549 | 5,492 |
| Adjusted $R^2$ | 0.0177 | 0.0049 | 0.0172 | 0.0152 | 0.0182 | 0.0272 | 0.0042 | 0.0108 |
| RMSE | 0.1782 | 0.2500 | 0.1766 | 0.1785 | 0.1782 | 0.1709 | 0.1773 | 0.1437 |
| AIC | -12999.25 | 737.27 | -13202.61 | -11747.32 | -11578.87 | -12093.41 | -12248.31 | -14411.19 |
| BIC | -12969.97 | 766.56 | -13173.32 | -11718.45 | -11550.48 | -12065.50 | -12220.10 | -14384.75 |

Note: (1) main specification, (2) random effects model, (3) gravity-type fixed effects model with 1% winsorizing by outflow rate data, (4) gravity-type fixed effects model with 5% trimming by outflow rate data, (5) gravity-type fixed effects model with random selected 80% samples of all samples, (6) gravity-type fixed effects model with excluding 2020 samples, (7) gravity-type fixed effects model with considering one-year lagged effects, (8) gravity-type fixed effects model with considering

*Corresponding author: Xing Su. Email address: 20250612@hhu.edu.cn

two-year lagged effects. Regression coefficients with standard errors in parentheses. We use cluster robust standard errors at the origin–destination pair level in all models. *P* values of coefficients were derived based on two-sided t-tests and no adjustments were made for multiple comparisons. *$P < 0.1$, **$P < 0.05$ and ***$P < 0.01$. The F-test is conducted between the gravity-type fixed effects model and the corresponding pooled OLS model and the Hausmann test is conducted between the gravity-type fixed effects model and the corresponding random effects model. The adjusted $R^2$, root mean square error (RMSE), Akaike information criterion (AIC) and Bayesian information criterion (BIC) evaluate the performance and goodness of fit of the models.

*Corresponding author: Xing Su. Email address: 20250612@hhu.edu.cn

**Table S5 Estimated marginal effects of HPs on migration under different socio-economic characteristics using interaction models with only linear interaction terms**

| | **Outcome: ln (annual out-migration rate)** | | | |
|---|---|---|---|---|
| | **(1)** | **(2)** | **(3)** | **(4)** |
| Institutional MP (origin) | -0.0820** | -0.0012 | -0.0026 | -0.0075* |
| | (0.0357) | (0.0046) | (0.0047) | (0.0038) |
| Behavioral/Cultural MP (origin) | 0.0591 | -0.0011 | 0.0103 | 0.0058* |
| | (0.0462) | (0.0037) | (0.0072) | (0.0033) |
| Behavioral/Cultural AP (destination) | -0.0732 | 0.0287** | 0.0130 | -0.0298* |
| | (0.1905) | (0.0142) | (0.0131) | (0.0160) |
| Technological/Infrastructural MP (destination) | 0.0560** | 0.0063* | 0.0008 | 0.0036 |
| | (0.0268) | (0.0034) | (0.0045) | (0.0025) |
| Behavioral/Cultural MP (destination) | -0.0042 | 0.0078** | -0.0075 | -0.0007 |
| | (0.0413) | (0.0032) | (0.0064) | (0.0032) |
| Nature-based MP (destination) | -0.1193 | -0.0579*** | -0.0241 | 0.0082 |
| | (0.1936) | (0.0179) | (0.0277) | (0.0170) |
| Institutional MP × moderator variable (origin) | 0.0074** | 0.0084 | 0.0489 | 0.0264*** |
| | (0.0031) | (0.0070) | (0.0345) | (0.0088) |
| Behavioral/Cultural MP × moderator variable (origin) | -0.0048 | 0.0097 | -0.0427 | -0.0018 |
| | (0.0041) | (0.0059) | (0.0540) | (0.0074) |
| Behavioral/Cultural AP × moderator variable (destination) | 0.0072 | -0.0431 | -0.0313 | 0.0986** |
| | (0.0168) | (0.0283) | (0.0835) | (0.0420) |
| Technological/Infrastructural MP × moderator variable (destination) | -0.0051** | -0.0135*** | -0.0234 | -0.0131** |
| | (0.0023) | (0.0049) | (0.0325) | (0.0055) |
| Behavioral/Cultural MP × moderator variable (destination) | 0.0001 | -0.0188*** | 0.0263 | -0.0075 |
| | (0.0036) | (0.0053) | (0.0482) | (0.0072) |
| Nature-based MP × moderator variable (destination) | 0.0110 | 0.1147*** | 0.2414 | -0.0045 |
| | (0.0171) | (0.0324) | (0.2063) | (0.0412) |
| Moderator variable (origin) | 0.8286*** | -0.8173*** | 2.5004 | 1.6623** |
| | (0.1796) | (0.2065) | (2.0782) | (0.7562) |
| Moderator variable (destination) | -0.7160*** | 1.2444*** | 0.7701 | -0.4039 |
| | (0.1920) | (0.2280) | (2.1305) | (0.7050) |
| Observations | 11,177 | 11,177 | 11,177 | 11,177 |
| Adjusted $R^2$ | 0.0383 | 0.0484 | 0.0240 | 0.0317 |
| RMSE | 0.1764 | 0.1755 | 0.1777 | 0.1770 |
| AIC | -13225.5122 | -13343.4402 | -13060.4640 | -13148.5923 |
| BIC | -13115.6880 | -13233.6160 | -12950.6398 | -13038.7681 |

*Corresponding author: Xing Su. Email address: 20250612@hhu.edu.cn

Note: (1) The interaction model using ln (median income) as the moderator variable, (2) The interaction model using racial diversity as the moderator variable, (3) The interaction model using ageing rate as the moderator variable, (4) The interaction model using education attainment rate as the moderator variable. Regression coefficients with standard errors in parentheses. We use cluster robust standard errors at the origin–destination pair level in all models. *P* values of coefficients were derived based on two-sided t-tests and no adjustments were made for multiple comparisons. *$P$ < 0.1, **$P$ < 0.05 and ***$P$ < 0.01. The adjusted $R^2$, root mean square error (RMSE), Akaike information criterion (AIC) and Bayesian information criterion (BIC) evaluate the performance and goodness of fit of the models.

*Corresponding author: Xing Su. Email address: 20250612@hhu.edu.cn

**Table S6 Estimated marginal effects of HPs on migration under different socio-economic characteristics using interaction models with both linear interaction terms and quadratic interaction terms**

| | **Outcome: ln (annual out-migration rate)** | | | |
|---|---|---|---|---|
| | **(1)** | **(2)** | **(3)** | **(4)** |
| Institutional MP (origin) | -5.4065***<br>(1.5487) | -0.0023<br>(0.0122) | -0.0984***<br>(0.0177) | 0.0790***<br>(0.0178) |
| Behavioral/Cultural MP (origin) | 1.6875<br>(2.7237) | -0.0065<br>(0.0073) | 0.1736***<br>(0.0364) | -0.0034<br>(0.0216) |
| Behavioral/Cultural AP (destination) | -0.3057<br>(7.4098) | 0.1798***<br>(0.0535) | 0.1237**<br>(0.0594) | 0.0480<br>(0.1157) |
| Technological/Infrastructural MP (destination) | 1.6336<br>(1.2420) | -0.0148*<br>(0.0079) | 0.1080***<br>(0.0160) | -0.0074<br>(0.0126) |
| Behavioral/Cultural MP (destination) | -5.4010**<br>(2.5521) | 0.0107*<br>(0.0062) | -0.0628**<br>(0.0309) | -0.0375*<br>(0.0203) |
| Nature-based MP (destination) | 12.3034<br>(9.1829) | -0.0532<br>(0.0479) | -0.5300***<br>(0.1154) | 0.0131<br>(0.0971) |
| Institutional MP × moderator variable (origin) | 0.9458***<br>(0.2725) | 0.0126<br>(0.0430) | 1.2416***<br>(0.2186) | -0.3876***<br>(0.0859) |
| Institutional MP × moderator variable$^2$ (origin) | -0.0413***<br>(0.0120) | -0.0037<br>(0.0376) | -3.4720***<br>(0.6410) | 0.4687***<br>(0.0998) |
| Behavioral/Cultural MP × moderator variable (origin) | -0.2902<br>(0.4785) | 0.0398<br>(0.0329) | -2.2460***<br>(0.5066) | 0.0330<br>(0.1034) |
| Behavioral/Cultural MP × moderator variable$^2$ (origin) | 0.0125<br>(0.0210) | -0.0334<br>(0.0347) | 7.2274***<br>(1.7483) | -0.0293<br>(0.1153) |
| Behavioral/Cultural AP × moderator variable (destination) | 0.0515<br>(1.2961) | -0.6789***<br>(0.2028) | -1.3930**<br>(0.7041) | -0.2908<br>(0.5697) |
| Behavioral/Cultural AP × moderator variable$^2$ (destination) | -0.0021<br>(0.0567) | 0.6418***<br>(0.1883) | 3.9767**<br>(2.0125) | 0.4590<br>(0.6797) |
| Technological/Infrastructural MP × moderator variable (destination) | -0.2850<br>(0.2175) | 0.0712**<br>(0.0298) | -1.4003***<br>(0.2065) | 0.0334<br>(0.0601) |
| Technological/Infrastructural MP × moderator variable$^2$ (destination) | 0.0124<br>(0.0095) | -0.0755***<br>(0.0279) | 4.2739***<br>(0.6622) | -0.0447<br>(0.0686) |
| Behavioral/Cultural MP × moderator variable (destination) | 0.9420**<br>(0.4471) | -0.0245<br>(0.0311) | 0.8002*<br>(0.4441) | 0.1691*<br>(0.0973) |

*Corresponding author: Xing Su. Email address: 20250612@hhu.edu.cn

| | | | | |
|---|---|---|---|---|
| Behavioral/Cultural MP × moderator variable$^2$ (destination) | -0.0411** (0.0196) | -0.0047 (0.0349) | -2.6361* (1.5895) | -0.1968* (0.1085) |
| Nature-based MP × moderator variable (destination) | -2.1285 (1.6123) | 0.1480 (0.2189) | 7.4364*** (1.6653) | -0.0030 (0.4742) |
| Nature-based MP × moderator variable$^2$ (destination) | 0.0920 (0.0707) | -0.0789 (0.2367) | -24.8765*** (5.9846) | -0.0544 (0.5556) |
| Moderator variable (origin) | -9.3435*** (2.5458) | -1.2140* (0.7285) | -0.2494 (5.2004) | 5.3168*** (1.6891) |
| Moderator variable (destination) | 15.6792*** (2.7067) | 3.6145*** (0.7570) | 3.3950 (6.1311) | 4.2928** (1.8388) |
| Moderator variable$^2$ (origin) | 0.4539*** (0.1128) | 0.4073 (0.7371) | 7.6336 (14.8000) | -3.4422** (1.6147) |
| Moderator variable$^2$ (destination) | -0.7182*** (0.1178) | -2.5456*** (0.7736) | -8.6203 (15.9579) | -4.9143*** (1.7033) |
| Observations | 11,177 | 11,177 | 11,177 | 11,177 |
| Adjusted $R^2$ | 0.0564 | 0.0579 | 0.0497 | 0.0462 |
| RMSE | 0.1748 | 0.1746 | 0.1754 | 0.1757 |
| AIC | -13429.3069 | -13447.6520 | -13351.0517 | -13309.7242 |
| BIC | -13260.9098 | -13279.2549 | -13182.6546 | -13141.3271 |

Note: (1) The interaction model using ln (median income) as the moderator variable, (2) The interaction model using racial diversity as the moderator variable, (3) The interaction model using ageing rate as the moderator variable, (4) The interaction model using education attainment rate as the moderator variable. Regression coefficients with standard errors in parentheses. We use cluster robust standard errors at the origin–destination pair level in all models. *P* values of coefficients were derived based on two-sided t-tests and no adjustments were made for multiple comparisons. **P* < 0.1, ***P* < 0.05 and ****P* < 0.01. The adjusted $R^2$, root mean square error (RMSE), Akaike information criterion (AIC) and Bayesian information criterion (BIC) evaluate the performance and goodness of fit of the models.

*Corresponding author: Xing Su. Email address: 20250612@hhu.edu.cn

## 2.2. Figures S1 to S5

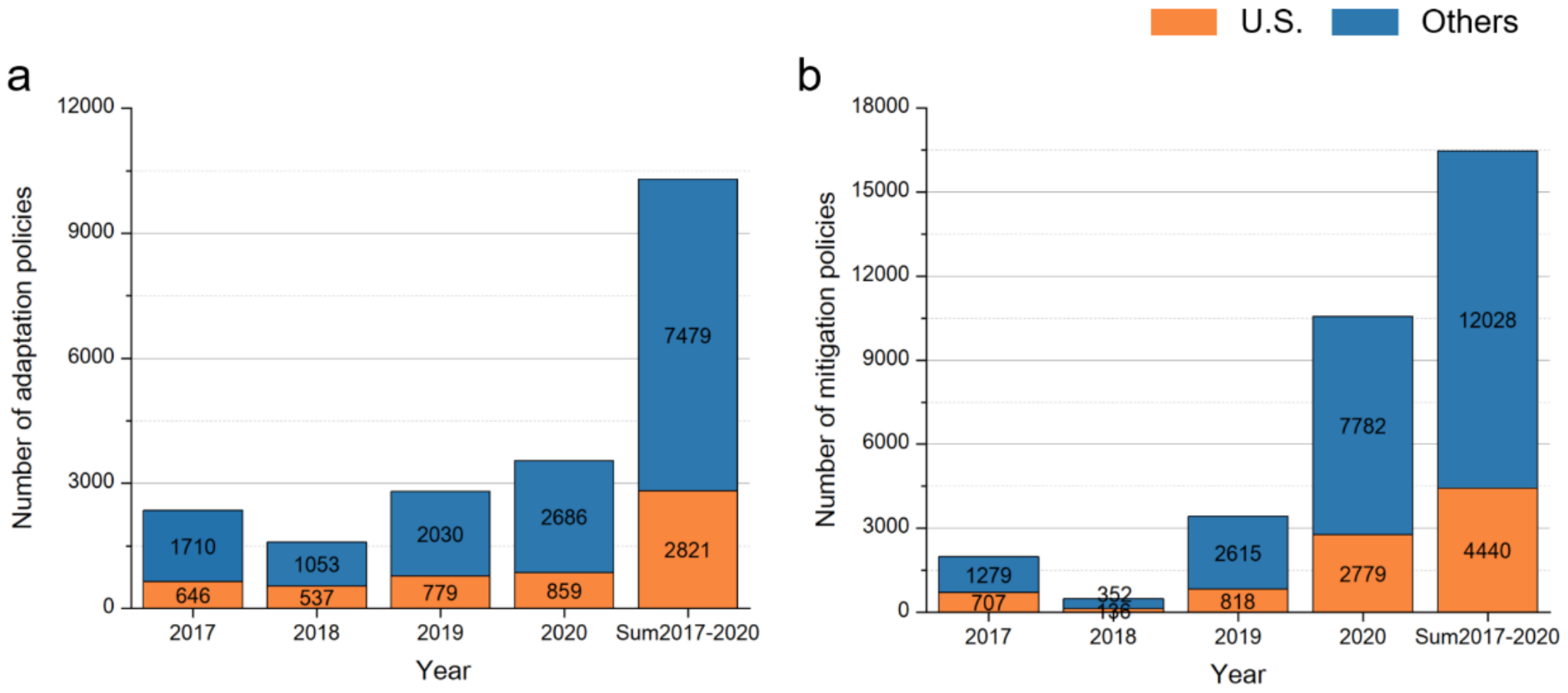

**Fig. S1 Number of climate policies in the U.S. and the rest of the world excluding the U.S. recorded by CDP during 2017-2020. a**, Number of adaptation policies. **b**, Number of mitigation policies.

*Corresponding author: Xing Su. Email address: 20250612@hhu.edu.cn

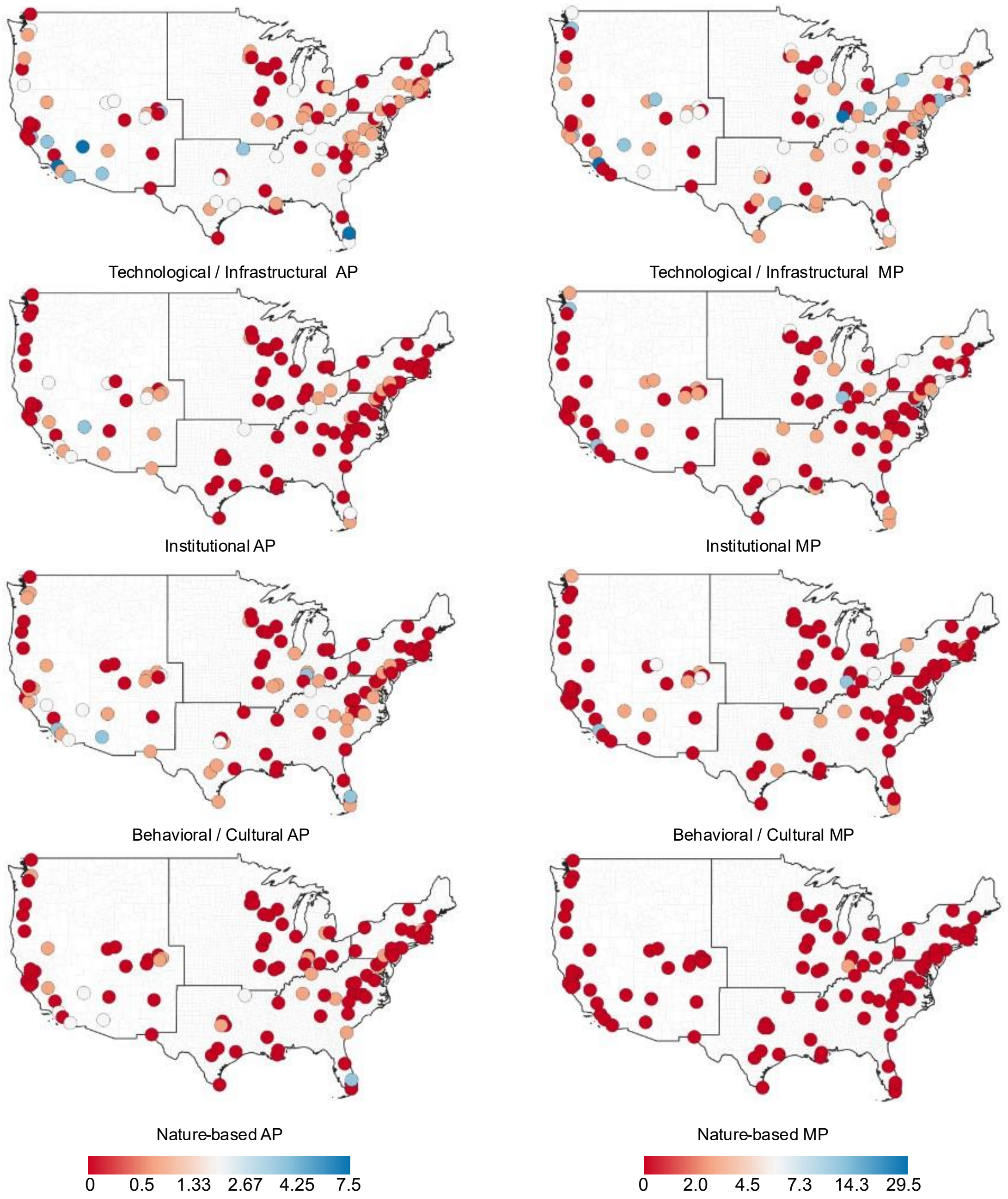

**Fig. S2 Geographic distribution of different policies at origin counties in 2017-2020.** The color of the circles represents the average number of different HPs implemented over the 2017-2020 period. Gray solid lines are U.S. County boundary lines, and the base map is from geodatabases of 2019 TIGER/Line shapefiles for US Geographical boundaries at county level.

*Corresponding author: Xing Su. Email address: 20250612@hhu.edu.cn

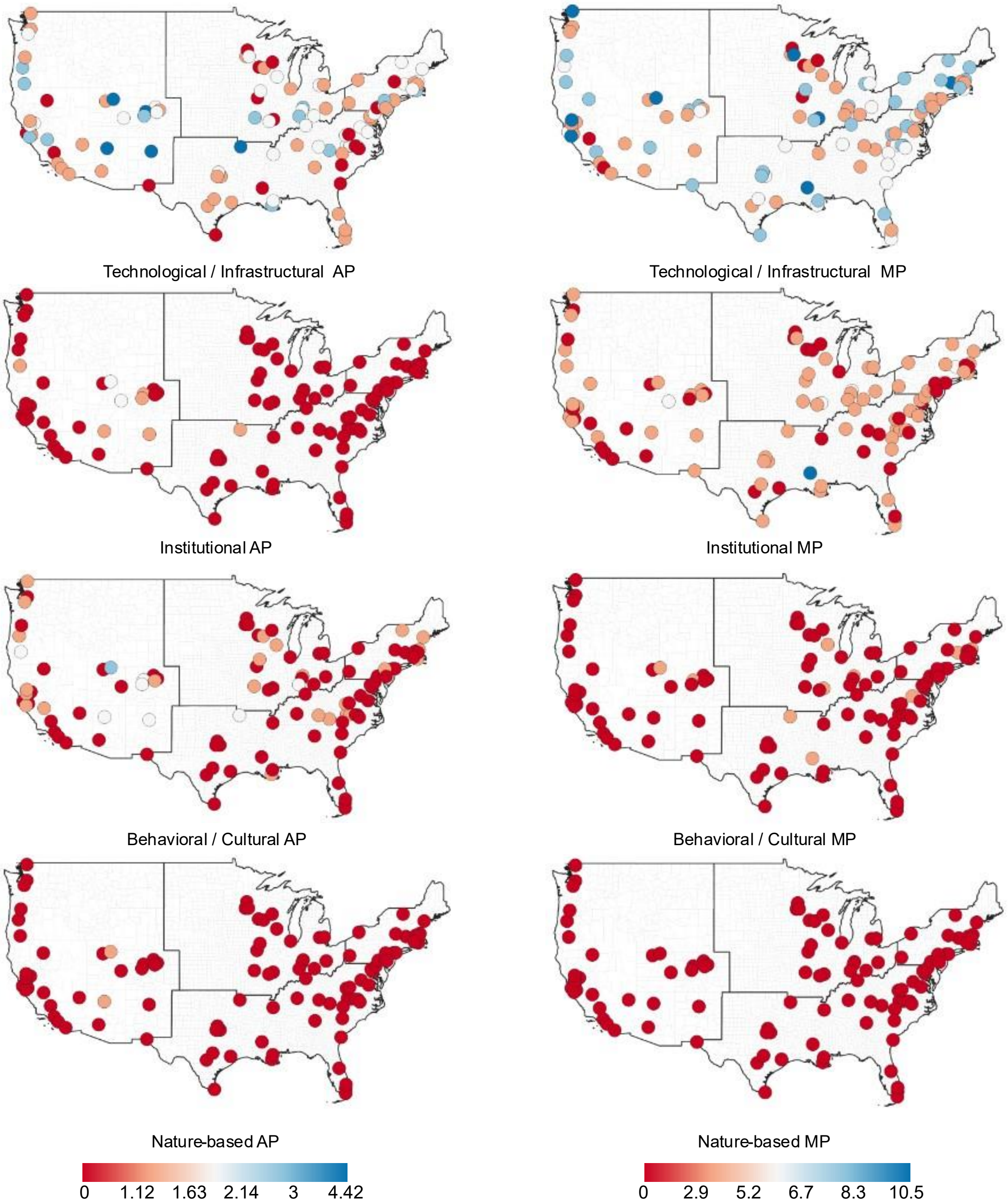

**Fig. S3 Geographic distribution of different policies at destination counties in 2017-2020.** The color of the circles represents the average number of different HPs implemented over the 2017-2020 period. Gray solid lines are U.S. County boundary lines, and the base map is from geodatabases of 2019 TIGER/Line shapefiles for US Geographical boundaries at county level.

*Corresponding author: Xing Su. Email address: 20250612@hhu.edu.cn

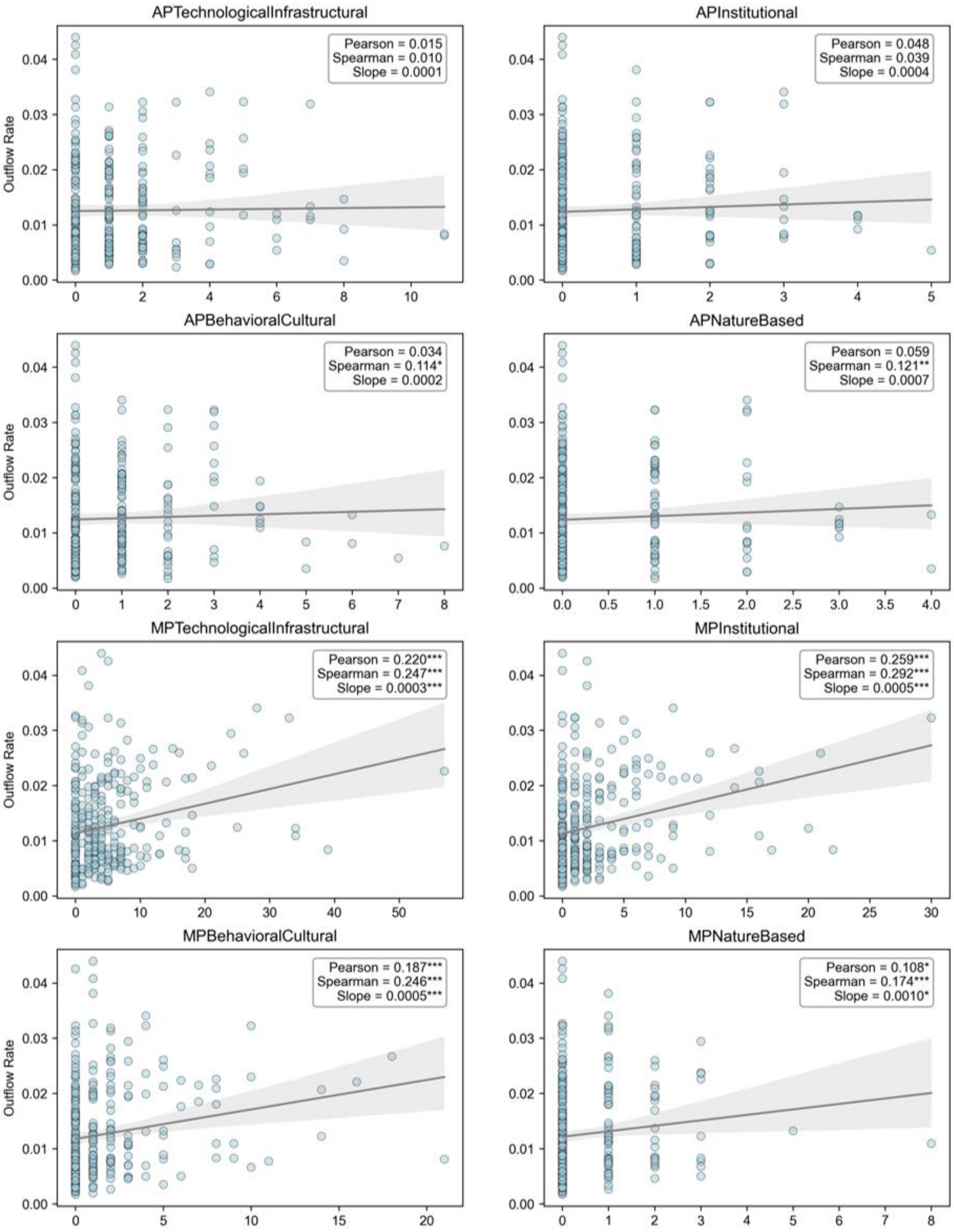

**Fig. S4 Scatterplots between different policy types implemented at origin counties and out-migration rate of origin counties.** The solid gray line indicates the line of fit for the simple linear regression, and the shaded area indicates the 95% confidence interval. The Pearson in the figures indicate the Pearson coefficients, the Spearman in the figures indicate the Spearman's rank coefficients, Slope indicates the coefficients of the linear regression. The symbol in the upper right corner of the figure indicates the significant $P$ values obtained from two-sided t-tests for that coefficient. *$P < 0.1$, **$P < 0.05$ and ***$P < 0.01$.

*Corresponding author: Xing Su. Email address: 20250612@hhu.edu.cn

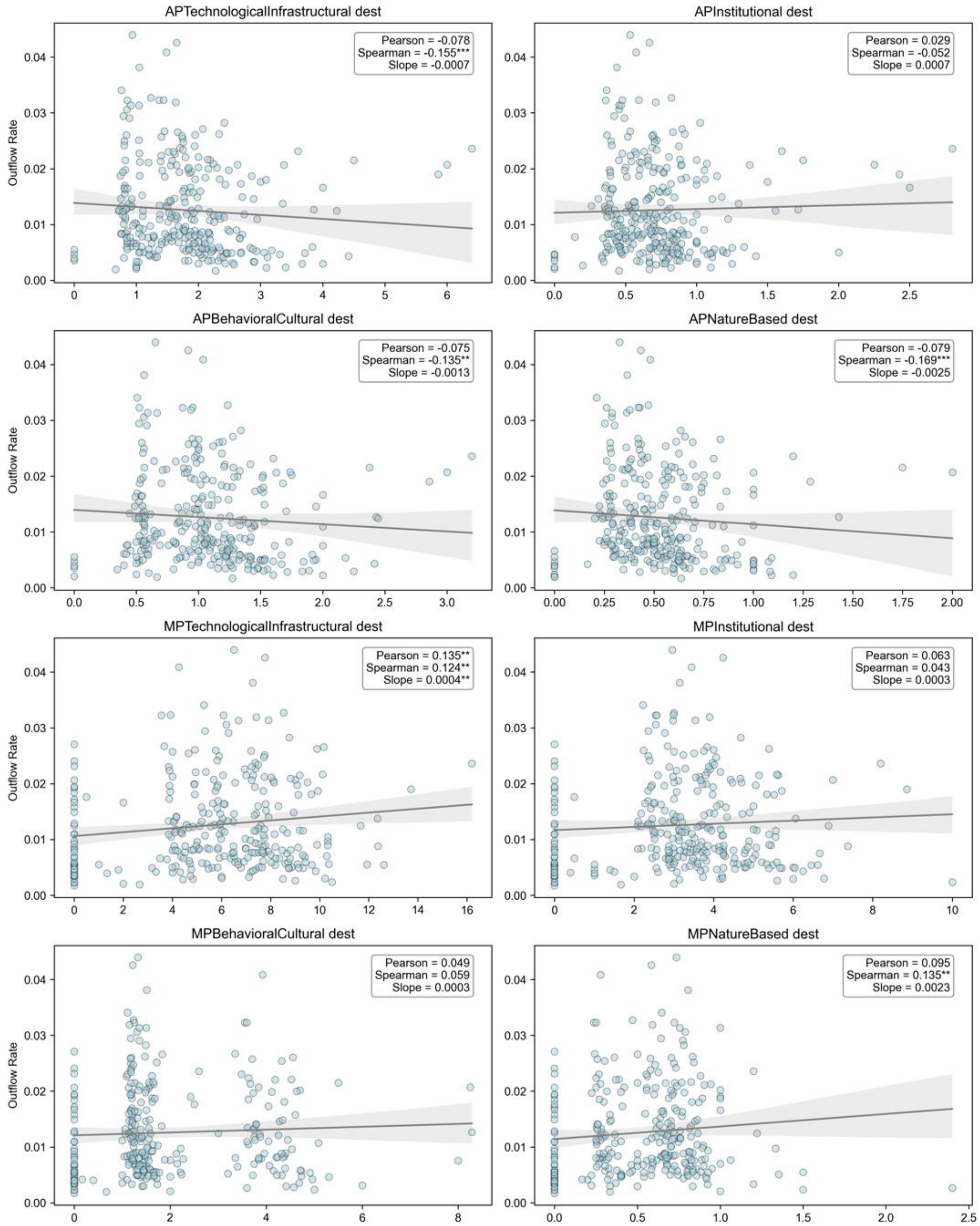

**Fig. S5 Scatterplots between different policy types implemented at destination counties and out-migration rate of origin counties.** The solid gray line indicates the line of fit for the simple linear regression, and the shaded area indicates the 95% confidence interval. The Pearson in the figures indicate the Pearson coefficients, the Spearman in the figures indicate the Spearman's rank coefficients, Slope indicates the coefficients of the linear regression. The symbol in the upper right corner of the figure indicates the significant $P$ values obtained from two-sided t-tests for that coefficient. *$P < 0.1$, **$P < 0.05$ and ***$P < 0.01$.

*Corresponding author: Xing Su. Email address: 20250612@hhu.edu.cn

**3.References**

*Corresponding author: Xing Su. Email address: 20250612@hhu.edu.cn